\documentclass[12pt]{article}
\usepackage{epsf}
\usepackage{graphicx}
\usepackage{amssymb}

\def\di{\displaystyle}

\def\bg{\begin{eqnarray}\begin{array}{rcl}\displaystyle}
\def\eg{\end{array} &\di    &\di   \end{eqnarray}}
\def\bm#1{\begin{eqnarray}\begin{array}{#1}\di}
\def\bmo#1{\begin{eqnarray*}\begin{array}{#1}\di}
\def\bml#1#2{\begin{eqnarray}\begin{array}{#1}\label{#2}\di}
\def\bgo{\begin{eqnarray*}\begin{array}{rcl}\displaystyle}
\def\ego{\end{array} &\di    &\di \nonumber  \end{eqnarray*}}

\def\btensor#1#2{\renew\left#1\begin{array}{#2}\di}
\def\brtensor#1#2#3{\ren#3\left#1\begin{array}{#2}}
\def\botensor#1#2{\renew\left#1\begin{array}{#2}}
\def\etensor#1{\end{array}\right#1}

\def\eq#1{(\ref{#1})}

\def\tr{{\rm tr}}

\def\T{{\rm T}}
\def\id{1\!\mbox{l}}

\def\ov{\over}

\def\be{\begin{equation}}
\def\ee{\end{equation}}
\def\bea{\begin{eqnarray}}
\def\eea{\end{eqnarray}}
\def\hM{{\hat M}}

\def\CG{\mathcal G}

\def\h{\hbox{$\frac{1}{2}$}}
\def\hm{\hat M^\dagger}
\def\hmi{{\hm}{}^{-1}}
\def\tv{{\tilde{\cal V}}}
\def\tV{\tv}
\font\af=msbm12
\def\T{{\mbox{\af T}}}
\def\R{{\mbox{\af R}}}
\def\Z{{\mbox{\af Z}}}
\def\H{{\mbox{\af H}}}

\def\lo{\frac{L_0}{2\pi}}
\def\i{{\mbox{i}}}

\date{\today}

\def\ren#1{\renewcommand{\arraystretch}{#1}}

\def\renew{\renewcommand{\arraystretch}{1}}
\ren{1.5}

\begin{document}

\begin{titlepage}

%\parindent=12pt
%\baselineskip=20pt
%\textwidth 15 truecm
%\vsize=23 truecm
%\hoffset=0.7 truecm
%\noindent{May 2000} 
\begin{flushright}
DESY 00-080\\
DIAS-STP-00-11\\
UNITU--THEP--6/2000\\
 FSUJ-TPI-06/2000 \\
  hep-th/0005221\\    
      \end{flushright}
\par
\vskip .5 truecm
\large \centerline{\bf ADHM Construction of Instantons}
\large \centerline{\bf on the Torus} 
\par
\vskip 0.5 truecm
\normalsize
\begin{center}
{\bf C.~Ford} ${}^{a}$, 
{\bf J.~M.~Pawlowski} ${}^{b}$, 
{\bf T.~Tok} ${}^{c}$ and
{\bf A.~Wipf} ${}^{d}$
\\
\vskip 0.5 truecm

${}^a$\it{Deutsches Elektronen-Synchrotron DESY-Zeuthen\\
Platanenalle 6, D-15738 Zeuthen, Germany\\ 
{\tt \,ford@ifh.de}}
\vskip 0.3 truecm

${}^b$\it{Dublin Institute for Advanced Studies\\
10 Burlington Road, Dublin 4, Ireland\\ 
{\tt \,jmp@stp.dias.ie}}
\vskip 0.3 truecm

${}^c$\it{Institut f\"ur Theoretische Physik, Universit\"at T\"ubingen\\
Auf der Morgenstelle 14, D-72072 T\"ubingen, Germany\\ 
{\tt \,tok@alpha6.tphys.physik.uni-tuebingen.de}}
\vskip 0.3 truecm

${}^d$\it{Theoretisch Physikalisches Institut, Universit\"at Jena\\ 
Fr\"obelstieg 1,
D--07743 Jena,
Germany\\ {\tt \,wipf@tpi.uni-jena.de}}
\end{center}
\vskip 0.5 true cm
\normalsize
\begin{abstract} 
We apply the ADHM instanton 
construction to $SU(2)$ gauge theory on
$\T^n\times\R^{4-n}$
for $n=1,2,3,4$. To do this we regard instantons on
 $\T^n\times\R^{4-n}$ as periodic (modulo
gauge transformations) instantons on $\R^4$.
Since  the $\R^4$ topological charge
 of such instantons is infinite
the ADHM  algebra
takes place on an infinite dimensional linear space.
The ADHM matrix  $ M$  is related to
 a Weyl operator (with a self-dual background)
on the dual torus $\tilde\T^n$. We construct the Weyl operator
corresponding to the
 one-instantons on $\T^n\times\R^{4-n}$. 
In order to derive the self-dual potential on $\T^n\times\R^{4-n}$
it is necessary to solve a specific Weyl equation. This is a variant 
of the Nahm transformation. In the case $n=2$ (i.e. 
$\T^2\times\R^2$)  we essentially  have
 an Aharonov Bohm problem on $\tilde \T^2$.
In the one-instanton sector we find that the scale parameter,
$\lambda$, is bounded  above, $\lambda^2\tv<4\pi$, $\tv$
being the volume of the dual torus $\tilde\T^2$.

\end{abstract}

\centerline{\sl Keywords: \rm\, Instantons; ADHM construction; Nahm
transformation}

\end{titlepage}
 \baselineskip=20pt
\setcounter{equation}{0}
\section{ Introduction}

Instantons are self-dual solutions of the pure Yang-Mills equations
\cite{bpst}.
For the classical groups the complete set of instanton
solutions on $\R^4$ (and via stereographic projection $S^4$)
have been known for over twenty years.
Although even now  some important details remain obscure.
For example,
what is the metric on the $k$-instanton moduli space \cite{measure,man,os}
for $\R^4$ instantons?
This is an important ingredient in the instanton-theoretic
checks \cite{fp,dorey,yung} of the Seiberg-Witten results \cite{sw}
in ${\cal N}=2$
supersymmetric Yang Mills theory.
For other four manifolds even less is known.
A particularly important manifold is the four torus $\T^4$. 
Firstly, it is compact, thereby removing from the outset,
any infrared divergences.
Unlike other compact four manifolds (e.g. $S^4$ or $K3$)
the four torus retains translational invariance, and  is flat.
However, while $\T^4$ has all these attractive features
the only known explicit $\T^4$ instanton solutions
 are some reducible
constant curvature solutions
due to 't Hooft \cite{thooft1}.
 These exist only for special values of the periods
and can only represent singular points in the moduli space
of a given instanton sector.
The possibility that these constant curvature solutions
are the only instantons on $\T^4$ was ruled out a long time ago by
Taubes
\cite{taubes}.
However, using the Nahm transformation, it can be shown that there
exist no untwisted instantons with unit topological charge on $\T^4$
\cite{vanbaal,vbnpbs}.
The  work of Taubes established the existence of instantons
in  \sl all higher \rm topological charge sectors.
A similar pattern is followed by the $O(3)$ sigma model instantons on
$\T^2$ \cite{rr}. Here the one instanton sector is empty, and 
this corresponds to the statement that there are no elliptic functions
with a single simple pole in the fundamental torus.

How should one start to  look for instanton solutions on $\T^4$?
An obvious  approach would be to adapt to the torus the 
techniques developed in the late 1970's for the $\R^4$ problem.
Loosely speaking, we seek periodic versions of these ans\"atze,
since instantons on $\T^4$ can be viewed as periodic solutions\footnote{
They can only be periodic in a singular gauge.}
on $\R^4$.
The general solution to the instanton problem on $\R^4$
was provided by Atiyah, Drinfeld, Hitchin and Manin (ADHM)
\cite{adhm}.
This work reduces the problem of constructing instantons
on
$\R^4$ or $\hbox{S}^4$ to an exercise in algebra.
To construct an instanton with topological charge
 $k$ one must find a quaternionic
$(k+1)\times k$
matrix, $M$,  obeying certain non-linear reality conditions.
However,  while this construction is purely algebraic,
its structure is very much tied to the manifold $\R^4$ or $S^4$,
and it appears difficult to `make it periodic' in a simple way.
An important subclass of solutions is provided by the 
't Hooft ansatz \cite{thooftansatz,thooft,wilczek,jnr}.
This converts a (singular) positive solution of the Laplace
equation
into an $SU(2)$ instanton.
Since this is a linear equation, it seems that we simply have to
find a periodic solution of the Laplace equation
to construct an instanton on the torus.
However, it is not too difficult to show that it is impossible
to construct a positive solution of the Laplace equation on $\T^4$ with
acceptable
singularities (i.e. singularities which do not show up
in the Yang-Mills  action density).

In this paper we  render the ADHM construction periodic by
`brute force', in that we regard instantons on the torus
as a periodic lattice of instantons on $\R^4$.
We  start with ADHM data corresponding
to an infinite array  of instantons embedded in $\R^4$.
While our initial objective
was to extract the $\T^4$ instantons, 
we will see that the less ambitious target to
have periodicity in fewer than four directions
offers considerable technical simplification.
To that end
   we consider the application of the ADHM
method to $SU(2)$ Yang-Mills on $\T^n\times\R^{4-n}$ for $n=1,2,3,4$.
Although $\T^4$ has no one instanton solution,
$S^1\times\R^3$, $\T^2\times\R^2$ and $\T^3\times\R$
should have \cite{vbnpbs}.
Again the $O(3)$-sigma model provides a useful hint,
since while there are no one-instantons on $\T^2$, one-instanton solutions
have been constructed on $S^1\times\R$ \cite{mw}.
As the $\R^4$ topological charge of a $\T^n\times\R^{4-n}$
instanton is infinite  we have to deal with an infinite dimensional
$M$ matrix.
For the $k$-instanton problem on $\T^n\times \R^{4-n}$, $ M$
can be related to  a $U(k)$ Weyl operator 
on $\tilde \T^n$, $\tilde \T^n$ being the torus dual to $\T^n$.
This is a manifestation of the Nahm transformation \cite{nahm,cg}.

Recently this programme has been implemented
by Kraan and
van Baal 
in the one-instanton sector of $SU(N)$ gauge theory on $S^1\times \R^3$ 
\cite{kraan,kraan2}.
Equivalent results were derived independently
by Lee and Lu \cite{lee}.
These works revealed a vivid
 `monopole constituent'
picture of calorons (see also \cite{mon,fmptw,ftw,jahn}). 
There is however an important pitfall in this whole approach;
even if one has constructed a Weyl operator  on $\tilde \T^n$ via the ADHM
method
one must check that it actually leads to a well defined gauge
potential on $\T^n\times\R^{4-n}$.\footnote{ For $n=1$
 the procedure always leads to a well defined
 instanton.}  
Here we 
solve the ADHM constraints for the one instanton problem
on $\T^n\times\R^{4-n}$ and give particular solutions
for the two instanton case.
However, we are only able to explicitly check that
these \sl sometimes \rm\,
lead to a well defined gauge potential for
 $n=2$. 
This is because the technical task of solving the Weyl equation
on $\tilde \T^n$ becomes more involved for higher $n$.
We will see that  the $n=2$ case (i.e. $\T^2\times\R^2$)
 boils down to a
specific  Aharonov Bohm  problem 
\footnote{
To our knowledge the extensive literature on the AB problem
(see for example \cite{abprops,st,osl})
does not explicitly tackle this specific case.}
on $\tilde \T^2$.
A stringy interpretation of $\T^2\times\R^2$
instantons 
can be found in \cite{stringy}.
Our gauge potential on $\T^2\times\R^2$ is well defined only
if we apply certain constraints on the ADHM parameters.
In the one instanton sector there is an upper limit
on the scale parameter.
For our subclass of two instantons further constraints emerge.
The two `component' instantons must share a common
scale parameter which itself is bounded from above. Furthermore,
the relative
group orientation of the two instantons is constrained.

The outline of this paper is as follows.
In chapter~\ref{sec2} we briefly 
recall the standard ADHM construction on $\R^4$ 
and then explain in a general way how it can 
be `made periodic' in one or more directions.
In chapter~\ref{sec3} we solve the ADHM constraints for
the one-instanton problem on $\T^n\times\R^{4-n}$.
The associated Weyl operator on $\tilde \T^n$ is given explicitly
in terms of a specific Green's function for the Laplace operator on
$\tilde\T^n$. Then we specialise to $\T^2\times\R^2$, where 
the Weyl equations 
 seem to be more manageable than in the general case.
Finally in chapter~\ref{sec4} we discuss the two instanton problem. 
Some technical results 
are given in the appendices. 

During the writing up of this paper we became aware of some
related work by Jardim. In a series of papers
\cite{jardim,jardim2,jardim3}
a mathematically sophisticated analysis of the Nahm transformation
on $\T^2\times\R^2$ has been given.
 A somewhat more physical account can be found in
\cite{periodicmonopoles} where the Jardim formalism is applied
to periodic monopoles, i.e. instantons on $S^1\times\R^2$
so that the dual torus is $\tilde S^1\times \R$ instead
of $\tilde\T^2$.

\setcounter{equation}{0}
\section{ADHM construction}\label{sec2}

In this chapter we review the standard ADHM construction on $R^4$.
We then explain how the formalism can be extended to
$\T^n\times\R^{4-n}$.
This is a straightforward extension of the 
$S^1\times\R^3$ formalism.

\subsection{ADHM on $\R^4$}

Closely following the presentation of
Christ Weinberg and Stanton  \cite{cws} (see also
 \cite{corrigan})       we briefly recall the ADHM
construction. For simplicity we specialise to the gauge group
$SU(2)$.
We wish to construct a self-dual $SU(2)$ Yang-Mills field
$A_\mu(x)$ on $\R^4$ with topological charge or
instanton number
\bg
k=-\frac{1}{16\pi^2}\int_{\R^4}d^4x\,\hbox{tr}\left(
F_{\mu\nu}F_{\mu\nu}\right).\eg
Here the Yang-Mills field strength is
\bg
F_{\mu\nu}=\partial_\mu A_\nu-\partial_\nu A_\mu +[A_\mu,
A_\nu],\eg
and the gauge field $A_\mu$ can be viewed as a $2\times 2$
anti-Hermitian traceless matrix.
However, one can equally regard $A_\mu$
as being a purely imaginary \sl quaternion. \rm
Recall  that the space of quaternions $\H$ has
 four generators $\i_\mu=(1,
\hat i,\hat j,\hat k)$
where the $\hat i$, $\hat j$, $\hat k$ anticommute and
satisfy
\bg
\hat i^2=\hat j^2=\hat k^2=-1,\quad \hat i\hat j\hat k=-1.\eg
The transition back to the standard Pauli matrix language can be made via 
the identifications
$\hat i\leftrightarrow -i{\bf\sigma}_1$,
$\hat j\leftrightarrow -i{\bf \sigma}_2$,
$\hat k\leftrightarrow -i{\bf \sigma}_3$.
We will use $*$ to denote quaternionic conjugation
(i.e. $1^*=1$, $\hat i^*=-\hat i$, $\hat j^*=-\hat j$,
$\hat k^*=-\hat k$). In the following $\dagger$ should be understood
as the transpose of the quaternionic conjugate.

The recipe for constructing a self-dual $A_\mu$ with instanton number $k$
is as follows.
One simply has to construct a $k+1\times k$
quaternionic matrix
$M$ with the following properties:

i) the $k\times k$ matrix $ M^\dagger M$ is real.

ii) $M$ is linear in the  quaternion
$x\equiv x_0+x_1\hat i+x_2\hat j +x_3\hat k$
formed from the four Euclidean coordinates.

The corresponding anti-hermitian
self-dual gauge potential is given by
\bg
A_\mu(x)=N^\dagger(x)\partial_\mu
N(x),\eg
where $N(x)$ is a $k+1$ component column vector
satisfying
\bg\label{ncondition}
M^\dagger N=0,\hbox{ and }
N^\dagger N=1.\eg  
Without loss of generality one may assume
$M$ has the following form \cite{cws,corrigan}
\bg
M=\btensor{(}{c}
v\\\di
\hat M\etensor{)},
\eg
where
$v$ is 
  a $k$-component row
vector $v$ made up of $k$ constant quaternions
\bg v=(q_1\, q_2\, ...\,q_k).\eg
These quaternions encode the scales and $SU(2)$ group orientation
of the $k$ `component' instantons.
$\hat M$
is a $k\times k$ matrix  with the following `canonical'
form
\bg
\hat M_{ij}(x)=\delta_{ij}(y_i-x)+b_{ij}.\eg
  $b_{ij}$
is independent of $x$, symmetric and
 has no diagonal entries ($b_{ij}=0$ for $i=j$).
The reality of $M^\dagger M$ translates into the following
non-linear requirement on $b_{ij}$
\bg\label{cancon}
\frac{1}{2}
(q_i^*q_j-q_j^*q_i)+(y_i-y_j)^*b_{ij}+\frac{1}{2}
\sum_{l=1}^k\left(
b_{li}^*b_{lj}-b_{lj}^*b_{li}\right)=r_{ij},\eg
for some real $k\times k$ matrix $r$.
The $y_i$ can be interpreted as the quaternionic positions of the
instantons. 
One can immediately write down a column vector $N$ satisfying
(\ref{ncondition})
\bg
N=\btensor{(}{c}
\frac{u}{\sqrt{\rho}}\\\di
-\frac{1}{\sqrt{\rho}}\left(\hat M^\dagger\right)^{-1}v^\dagger\,
u\etensor{)},
\eg 
and
\bg\label{rho} \rho=1+v\hat M^{-1}\left(\hat M^\dagger\right)^{-1}v^\dagger.
\eg
Here  $u$ is an arbitrary, possibly $x$-dependent unit quaternion;
different choices for $u$ yield gauge equivalent Yang-Mills
fields.
Observe that it is necessary to invert the canonical form
$\hat M$ to extract the final gauge potential.
In the singular gauge $u(x)=1$, the potential can be written,
\bg\label{pot}
A_\mu=-{1\over{2\rho}}v\left(
\hM^{-1}\partial_{\mu}\hmi-\partial_\mu(\hM^{-1})\hmi
\right)v^\dagger.\eg
The corresponding field strength reads
\bg
F_{\mu\nu}=
\frac{1}{\rho}
v\hat M^{-1}\i_\mu~ f~ \i_\nu^*
(\hat M^\dagger)^{-1}v^\dagger-[\mu\leftrightarrow \nu],\eg
where
$f$ is the \sl real \rm $k\times k$ matrix
\bg
f=(M^\dagger M)^{-1}=
\hat M^{-1}(\hat M^\dagger)^{-1}-
\frac{1}{\rho}
\hat M^{-1}(\hat M^\dagger)^{-1}v^\dagger v\hat M^{-1}
(\hat M^\dagger)^{-1}.
\eg
The reality of $f$ ensures that 
$F_{\mu\nu}$ is self-dual.

One immediately sees that $A_\mu(x)$ is unaffected by the following
transformation on the ADHM data
\bg\label{orthogonal}
\hat M\rightarrow O^{-1}\hat M O,\quad
v\rightarrow v O,\eg
where $O$ is a $k\times k$ real orthogonal matrix.
Invoking this freedom one may argue  that $r_{ij}$ can be set to
zero \cite{cws}.
With this choice $b_{ij}$ is fully determined by the $8k$
parameters encoded in the $q_i$ and $y_i$.
Three of these parameters correspond to the global gauge symmetry.
This freedom can be fixed by taking $q_1$ to be real,
leaving $8k-3$ genuine moduli parameters.
A trivial but useful
 consequence of the `symmetry'
(\ref{orthogonal}) is that the $q_i$ are determined only up to
a sign.
If we flip the sign of one of the $q_i$, say $q_3\rightarrow-q_3$,
then this corresponds to the orthogonal transformation
$O=\hbox{diag}(1,1,-1,1,1,....)$.

\subsection{ ADHM on $\T^n\times\R^{4-n}$}\label{fourier}
 
We view $\T^n$ as $\R^n$ modulo a $n$ dimensional lattice $\Lambda$
generated by $n$ quaternions  $e_0$, $e_1$, ... ,$e_{n-1}$
corresponding to $n$ orthogonal vectors.
The periods or equivalently the Euclidean lengths of the $e_i$ are
denoted by $L_i,\quad i=0,1,...,n-1$.  
First we will show  how 
(in principle) one can produce instantons which in the singular gauge
(i.e. $u(x)=1$ as in eqn. (\ref{pot})) are periodic with respect
to shifts by the lattice generators, 
\bg\label{shifts}
A_\mu(x+e_i)=A_\mu(x),\quad i=0,1,..,n-1.\eg
Later we will consider a more general periodicity property which proved
crucial in obtaining new 1-instanton solutions on $S^1\times \R^3$.
To construct a k-instanton on $\T^n\times \R^{4-n}\equiv
\R^4/\Lambda$ consider the
following set up.
For every $\alpha\in \Lambda$ we have instantons at the positions
$y_i+\alpha$ with respective scale/orientation quaternions
$q_i$ where $i=1,2,...,k$ enumerates the instantons in the fundamental cell.
The quaternions $y_i$ give the instanton positions in the fundamental
cell.
Thus, our $\hat M$ and $v$ now have the following structure
\bg\label{periodic}
v_i^\alpha=q_i,\quad
{\hat M}_{ij}^{\alpha\beta}=\delta_{ij}\delta^{\alpha\beta}(y_i+\alpha
-x)
+b_{ij}^{\alpha\beta},\quad i,j=1,2,...,k,\quad
 \alpha,\beta\in \Lambda.\eg
The matrix $b_{ij}^{\alpha\beta}$ has the  properties
\bg
b_{ij}^{\alpha\beta}=b_{ji}^{\beta\alpha},\quad 
b_{ii}^{\alpha\alpha}=0 \quad\hbox{ (no sum)},\eg
and
\bg\label{cancontor}
\frac{1}{2}
(v_i^\alpha{}^*v_j^\beta-v_j^\beta{}^*v_i^\alpha)
+(y_i-y_j+\alpha-\beta)^*b_{ij}^{\alpha\beta}
+\frac{1}{2}
\sum_{l=1}^k\sum_{\gamma\in\Lambda}\left(
b_{li}^{\gamma\alpha}{}^*b_{lj}^{\gamma\beta}
-b_{lj}^{\gamma\beta}{}^*b_{li}^{\gamma\alpha}
\right)=0.\eg
Now that $\hat M$ is an infinite dimensional matrix
the non-linear constraint appears much more formidable than 
its $\R^4$ counterpart (\ref{cancon}).
Moreover, even if we can solve the constraint we still face
the problem of inverting $\hat M$.
 We see  that the constraint implies $b^{\alpha \beta}_{ij}$
has the following property
\bg\label{required}
\hat b_{ij}^{\alpha \beta}=b_{ij}^{\alpha-\beta\,\,\, 0},\quad 
\alpha,\beta\in \Lambda.
\eg 
At this point it is useful to perform a Fourier transform \cite{kraan};
\bg
v_i(z)=\sum_{\alpha\in \Lambda}v_i^\alpha e^{- i\alpha\cdot z},\quad
\hat M_{ij}(z)\delta^{n}(z-z')=
\sum_{\alpha,\beta\in\Lambda}
\hat M_{ij}^{\alpha\beta}e^{i\alpha\cdot z
-
i\beta\cdot z'},\eg
where $\delta^{n}(z-z')$ is a $n$-dimensional delta function
which is periodic with respect to the \sl dual lattice \,\rm
 \bg\tilde
\Lambda=\{z\in\R^n|(2\pi)^{-1} z\cdot\alpha\in \Z \hbox{ for all }\alpha\in
\Lambda\}.\eg
Here $ \alpha \cdot z $ \sl denotes the usual scalar product in
$\R^n$,
\rm\, i.e.
$\alpha \cdot z = \sum_{j=0}^{n-1} \alpha_j z_j $.
The delta function has the Fourier representation
\bg
\delta^n(z)=\frac{1}{\tV} \sum_{\alpha\in\Lambda}e^{i\alpha\cdot z},\eg
where 
\bg\tv=(2\pi)^n/L_0 L_1 ... L_{n-1},\eg
is the volume of the \sl dual torus \rm\,
$\tilde\T^n:=\R^n/\tilde\Lambda$.
Using (\ref{periodic}) $\hat M_{ij}$ can be written as follows
\bml{c}{dz}
\tv^{-1}\hat M_{ij}(z)=\delta_{ij} \left( 
-i d_z-x+\frac{1}{k}\sum_{l=1}^k y_l \right) -i \hat A_{ij}(z),
\quad \quad  d_z=\sum_{i=0}^{n-1}\i_i{\partial_{ z_i}},\eg
and 
\bg 
-i\hat A_{ij}(z)=\delta_{ij} \left(y_i-\frac{1}{k}\sum_{l=1}^k y_l\right)
+\sum_{\alpha\in\Lambda}b_{ij}^{\alpha 0}
e^{i\alpha\cdot z},
\eg 
can be regarded as a $SU(k)$ ($U(1)$ for $k=1$)
potential on the dual torus
$\tilde \T^n$.
From now on we will assume (without loss of generality) that
\bg
\sum_{l=1}^ky_l=0,\eg
so that $\tv^{-1}\hat M_{ij}(z)=\delta_{ij}(-id_z-x)-i\hat A_{ij}(z)$.
The $z$-space analogue of $M$ can be written
as
\bg
M=\btensor{(}{c}
v_i(z')\\\di
\hat M_{ij}(z)\delta^n(z-z')\etensor{)}.
\eg
We also require $M^\dagger$
\bg
M^\dagger=
\btensor{(}{cc}
(v^\dagger)_i(z)&
(\hat M^\dagger )_{ij}(z)\delta^n(z-z')
\etensor{)},\eg
where 
\bg
(v^\dagger)_i(z)=\sum_{\alpha\in\Lambda}\left(
v_i^\alpha\right)^*e^{i\alpha\cdot z},\quad
(\hat M^\dagger)_{ij}(z)\delta^n(z-z')=
\sum_{\alpha,\beta\in\Lambda}
\left(M_{ji}^{\beta\alpha}\right)^*e^{i\alpha\cdot z -i\beta\cdot z'},
\eg
so that $\tv^{-1}\hat M_{ij}^{\dagger}(z)=\delta_{ij}(-id_z^*-x^*)-i
\hat A_{ij}^*(z)$.
We now consider the product $M^\dagger M$
\begin{eqnarray}
(M^\dagger M)_{ij}(z,z')
&=&(v^\dagger)_i(z)v_j(z')+
\tv^{-1} \int_{\tilde \T^n}d^n w
(\hat M^\dagger)_{ik}(z)\delta^n(z-w)
\hat M_{kj}(w)\delta^n(w-z')\nonumber \\
&=&(v^\dagger)_i(z)v_j(z')\nonumber \\
&&+
\tv^{-2}\left(
\delta_{ik}(
-id_z^*-x^*)-i\hat A_{ik}^*(z)\right)
\left(\delta_{kj}(-id_z-x)-i\hat A_{kj}(z)\right)
\delta(z-z').\nonumber \\
&&\, \end{eqnarray}
In $z$-space the constraint that $M^\dagger M$ is real
reduces to the self-duality equation for the 
$SU(k)$ $\bigl($ or $U(1)$ $\bigr)$
 potential $\hat A_{ij}(z)$, 
but
with delta function sources. 
These sources come from  the $(v^\dagger)_i(z)v_j(z')$
term;
with the choice (\ref{periodic}) we have $v_i(z)=\tv q_i\delta^n(z)$.

It is also possible to arrange so that in the singular gauge $u(x)=1$,
$A_\mu(x)$ is periodic modulo global gauge transformations.
This is achieved by replacing $v_i^\alpha=
q_i$ with
\bg\label{rotation}
v_i^\alpha=e^{ (\alpha\cdot \omega) \hat l}q_i,\eg
where $\omega$ is an element of the dual torus and
$\hat l$ is a purely imaginary unit quaternion.
In the $u(x)=1$ gauge, the instanton potential  has the following
periodicity properties
\bg\label{moregeneral}
A_\mu(x+e_i)=e^{ (e_i\cdot \omega) \hat l}
A_\mu(x)e^{- (e_i\cdot \omega) \hat l}.\eg
This choice of $v_i^\alpha$ still entails delta function
sources on the dual torus 
\bg\label{vformula}
v_i(z)=\h\tv\left[\left(1-i\hat
l\right)\delta^n(z-\omega)+\left(1+i\hat l\right)
\delta^n(z+\omega)\right]q_i.\eg
 $\left(1+i\hat l\right)$ and $\left(1-i\hat l\right)
$ are \sl projectors \rm
in the sense that
\bg
\left(1\pm i\hat l\right)^2=2\left(1\pm i\hat l\right),\quad
\left(1+i\hat l\right)\left(1-i\hat l\right)=0.\eg

Looking at the expression (\ref{pot}) for the $\R^4$ gauge potential we see
that it suffices to compute the
$k$-component row vector $n:=v\hat M^{-1}$.
The $\T^n\times\R^{4-n}$ analogue of this object
is the $z$-dependent $k$-component row vector, $n(z)$,
with components
\bg
n_j(z)=\tv^{-1} \sum_{i} \int_{\tilde \T^n}
d^nz'\, v_i(z')\hat M_{i j}^{-1}(z',z),\eg
and similarly the $k$-component column vector $n^\dagger(z)$
has components 

\noindent{$(n^\dagger)_i(z)=
\tv^{-1}\sum_j \int_{\tilde\T^n}d^n z'
(\hat M^\dagger)^{-1}_{ij}(z,z')(v^\dagger)_j(z')$.
Here

\noindent$\hat M_{ij}^{-1}(z,z')=
\sum_{\alpha,\beta}\left(
\hat M^{-1}\right)_{ij}^{\alpha \beta}
e^{i\alpha\cdot z-i\beta\cdot z'}$, so that
\begin{equation}
\hat M(z)\hat M^{-1}(z,z')=\tv^2\delta^n(z-z').\end{equation}
Using (\ref{vformula}) we have}
\bg\label{finaln}
n_j(z)=\h\left(1-i\hat l\right) q_i \hat M^{-1}_{ij}(\omega,z)+
\h\left(1+i\hat l\right)q_i \hat M^{-1}_{ij}
(-\omega,z),\eg
which reduces to $n_j(z)
=q_i\hat M_{ij}^{-1}(0,z)$ in the periodic case
 ($\omega=0$).
The $\T^n\times\R^{4-n}$ gauge potential can be written
\bm{c}\label{amu}
A_\mu=-\frac{\tv^{-1}}{2\rho}\int_{\tilde \T^n}
d^n z\left[
n(z)\partial_\mu{n^\dagger(z)}-
\partial_\mu(n(z)){n^\dagger(z)}\right],\eg
where $\rho$ is now 
\bg\label{rho1}\label{integ}
\rho=1+\tv^{-1}\int_{\tilde \T^n} d^n z\, n(z){n^\dagger(z)}.\eg
Note that the \sl integrand, \rm\,
$n(z){n^\dagger(z)}$
in (\ref{rho1}) is \sl not necessarily real, \rm
 although the integral itself,
$\int d^n z\, n(z){n^\dagger(z)}$,
is real and positive (see section
~\ref{6}). 
 
The corresponding field strength is
\bg\label{fieldstrength}
F_{\mu\nu}=\frac{{\cal V}^{-2}}{\rho}
\int_{\tilde \T^n}d^nz
\int_{\tilde \T^n}d^nz'
n(z)\i_\mu
f(z,z')~{\i_\nu^*}{n^\dagger(z')}-
[\mu\leftrightarrow \nu],
\eg
where the Green's function $f(z,z')$ is
\begin{eqnarray}\label{greensfunctionf}
f(z,z')&=&(M^\dagger M)^{-1}(z,z')\\ \nonumber
&=&\tv^{-1}\int_{\tilde \T^n}d^ny
\hat M^{-1}(z,y){
(\hat M^\dagger)^{-1}(y,z')}\\ \nonumber
&&-
\frac{{\tv}^{-2}}{\rho}
\int_{\tilde\T^n}d^ny
\hat M^{-1}(z,y){n^\dagger}(y)
\int_{\tilde\T^n}d^ny'
n(y'){(\hat M^\dagger)^{-1}(y',z')
}.\end{eqnarray}
 As we shall see,
 all the formulae in this section
 require particularly careful handling for $n>1$.

\setcounter{equation}{0}
\section{One-instantons}\label{sec3}

In this chapter we consider in some detail the one instanton problem
on $\T^n\times\R^{4-n}$. In particular we explicitly determine the 
ADHM matrix $M$. Under the Fourier transform this becomes a Weyl
operator associated with an Abelian self-dual potential
$\hat A(z)$ on the dual torus $\tilde\T^n$.
Unfortunately we do not have a general approach to the solution of such
 Weyl equations.
In section 3.2 we concentrate our attention on the $\tilde \T^2$
Weyl equation (corresponding to one instantons on
$\T^2\times\R^2$)
where $\hat A(z)$ is an Aharonov Bohm potential on $\tilde\T^2$.
The ADHM construction of the instanton potential $A_\mu(x)$
and $F_{\mu\nu}(x)$ is considered.
For values of $x$ restricted to a two dimensional subspace of
$\T^2\times \R^2$ closed forms for $A_\mu(x)$ and $F_{\mu\nu}$
are given.
From a mathematical standpoint the calculation is not completely 
satisfactory;
a formal limiting procedure is employed to obtain
the gauge potential.
However, we are able to check that 
the field strength  is self-dual
and that $\tr (F_{\mu\nu})^2$ is non-zero and smooth.
Moreover, in section 3.3 we see that our potential can be interpreted as the
Nahm transform of the AB potential $\hat A(z)$.
More specifically,
we identify the two Nahm zero modes associated with $\hat A(z)$.

\subsection{ADHM constraints for $\T^n\times \R^{4-n}$}\label{sec31}

Let us start by considering $1$-instanton solutions on
$\T^n\times \R^{4-n}$.
If we seek instantons which are strictly periodic in the $u(x)=1$
gauge we are immediately restricted to $S^1\times\R^3$.
This is because all the instantons in our lattice will, by
construction, have the same scale/group orientation $q_1$ and hence
be of the 't Hooft type.
Since the 't Hooft instantons 
 on $S^1\times \R^3$ are well known
\cite{hs} we will examine the more general instanton array
(\ref{rotation}).

Without loss of generality we can assume that $q_1$ is a real
quaternion which we identify as the `scale' $\lambda$, so that
\bg
v^\alpha=e^{(\alpha\cdot\omega)\hat l}\lambda,\eg
where we have dropped the redundant $1$ subscript on $v^\alpha$.
The $\hat M$ matrix has the form
\bg
\hat M^{\alpha\beta}=\delta^{\alpha\beta}(\alpha-x)+b^{\alpha\beta}.\eg
We now have to determine the $b$ matrix via (\ref{cancontor}).
Under the Fourier transformation this is a self-duality equation on the
dual torus $\tilde \T^n$. However, it is instructive to examine the
constraint equation in the original (matrix) variables.
In Appendix A we will argue that for $k=1$ the 
quadratic term in (\ref{cancontor})
is zero, i.e.
 the $b$ matrix
is simply
\bg\label{bmatrix}
b^{\alpha\beta}=-\frac{1}{2(\alpha-\beta)^*}
\left({v^\alpha}^*v^\beta-{v^\beta}^*v^\alpha\right)
=\frac{\lambda^2}{(\alpha-\beta)^*}
\hat l\, \sin\left[(\alpha-\beta)\cdot\omega\right],\quad
\alpha\neq\beta.
\eg
 In order to construct the potential we must now invert the
$\hat M$ matrix. To facilitate this we perform the Fourier transform
elaborated in section~\ref{fourier}, 
\bg
\tv^{-1}\hat M(z)=-i d_z-x-i\hat A(z),\eg
where $\hat A(z)$ is the $U(1)$ potential
\bg
\hat A(z)=i\lambda^2 d_z\phi(z)\hat l,\eg
and $\phi$ is the real function
\bg\label{phi}
\phi(z)=-\frac{1}{2}\sum_{\alpha\in\Lambda\setminus 0}
\frac{\exp[i\alpha\cdot(z+\omega)]-
\exp[i\alpha\cdot(z-\omega)]}{|\alpha|^2},\eg
which is a Green's function for the Laplace operator
on $\tilde\T^n$
\bg
 d_z d_z^*\phi(z)=\frac{\tv}{2}\left[
\delta^n(z+\omega)-\delta^n(z-\omega)\right].\eg
Clearly $\phi(z)$ is an odd function
\begin{equation}\label{odd}
\phi(-z)=-\phi(z).\end{equation}
Writing $\hat A(z)=\sum_{l=0}^{n-1}\i_l \hat A_l(z)$,
one can check that the Abelian field strength $\hat F_{ij}(z)=
\partial_i \hat A_j-\partial_j \hat A_i$ is self-dual, except at the
singularities
$z=\pm\omega$.

\subsection{One-instantons on $\T^2\times\R^2$}\label{6}

%In the special case $n=2$ (i.e. 
% $\T^2\times\R^2$) the Weyl operators are equivalent to (massive)
%Dirac operators for an Aharonov-Bohm problem on $\tilde \T^2$.
%The Dirac  mass roughly corresponds
%to how far the point $x\in \T^2\times\R^2$ is from the two-dimensional
%plane in $\T^2\times\R^2\subset \R^4$ where the instanton lattice lies.

Since our lattice is two dimensional we may take $e_0$ to be
real
and $e_1$ to be proportional to the purely imaginary unit quaternion
$\hat l$ \footnote{
We can always perform an $O(4)$ Lorentz transformation to arrange this.}.
Now rewrite the quaternion $z$ as follows
\begin{equation}\label{complexcoords}
z=z_0+\hat l z_1=
\h\left(1-i\hat l\right)\mbox{z}+\h\left(1+i\hat l\right)
\bar{\mbox{z}},
\end{equation}
where $\mbox{z}=z_0+iz_1$, $\bar{\mbox{z}}=z_0-iz_1$ denote standard
complex coordinates.
We can write the Fourier transformed $\hat M$ as follows
\bg\label{hatm}
\tv^{-1}\hat M(z)=-i d_z-x- i\hat A_0(z)-i\hat l\hat A_1(z),\eg
where
\bg\label{abpotential}
\hat A_0=-i\lambda^2\partial_{z_1}\phi,\quad
\hat A_1=i\lambda^2\partial_{z_0}\phi,\eg
and $\phi$ is the Green's function defined by (\ref{phi}).
Since we are on $\tilde\T^2$ we can write $\phi$ directly
in terms of Jacobi theta functions\footnote{
We follow the notation of Mumford \cite{mumford};
$\theta(\mbox{z},\tau)=\sum_{n=-\infty}^\infty
e^{ \pi in^2\tau+2 \pi in\mbox{z}}$.
In the fundamental torus $\theta(z,\tau)$ has a single
 zero
at $z=\h+\h\tau$, and has the periodicity properties
$\theta(\mbox{z}+1,\tau)=\theta(\mbox{z},\tau),\quad
\theta(\mbox{z}+\tau,\tau)=e^{- \pi i\tau-2\pi i \mbox{z}}\theta(
\mbox{z},\tau)$.}
\bg
\phi(\mbox{z})&=&\di\!\! \frac{\tv}{8\pi}\log
\frac{\left|\theta\left(\lo(\mbox{z}+\mbox{w}) +\h+\frac{iL_0}{2L_1}
,\frac{iL_0}{L_1}\right)
\right|^2}
{\left|\theta\left(\lo( \mbox{z}-\mbox{w})+\h+
\frac{iL_0}{2L_1}
,\frac{iL_0}{L_1}
\right)\right|^2
}+
\frac{(\mbox{z}-\bar{\mbox{z}})(\mbox{w}-\bar{\mbox{w}})}{4}-
i {{\rm w}-\bar {\rm w}\ov 4 L_1}
,\eg
where $\mbox{w}=\omega_0+i\omega_1$, $\bar{\mbox{w}}=\omega_0-i
\omega_1$.
The associated field strength is given by
$\hat F_{01}=i\lambda^2\square\phi$, which is zero except at $z=\pm \omega$.
At the points $\omega+\tilde\alpha,\quad
\tilde\alpha\in\tilde\Lambda$ we
have a `flux tube'
of strength $\h \lambda^2\tv$, and at
the points $-\omega+\tilde\alpha,\quad
\tilde\alpha\in\tilde\Lambda$
we have flux tubes of strength $- \h \lambda^2\tv$.

\begin{figure}[ht]
\begin{minipage}[ht]{16cm}
\centerline{\epsfysize=6 cm\epsffile{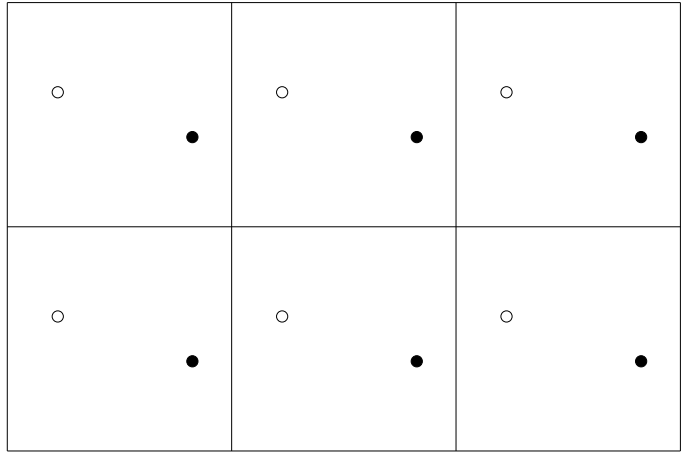}}
\caption{\label{fluxtubes}\textsl{
Flux tubes threading the dual torus at the points 
$\omega+\tilde\alpha$ and $-\omega+\tilde\alpha$ with equal and opposite
strengths.}}
\end{minipage} \end{figure}

What about the $x$ term in (\ref{hatm})?
It will prove convenient to decompose
$x$ into two pieces
\bg
x=x_{||}+x_\perp,\eg
where $x_{||}$ and $x_\perp$
respectively commute and anticommute with $\hat l$. 
Therefore the $x_{||}$ contribution
just amounts to shifting $\hat A_0$ and $\hat A_1$ by constants, while
$x_\perp$ is akin to a mass term.

We can write $\hat M(z)$ as follows
\bg\label{rewrite}
\tv^{-1}\hat M(z)=e^{-i \hat l\lambda^2\phi(z)} \left(-i d_z-x_{||}
\right)e^{i\hat l\lambda^2\phi(z)}-x_\perp.\eg
This is \sl not \rm a pure gauge decomposition
since the argument of the exponential is not a pure phase. 
If $x_\perp=0$, one can immediately write down a formal inverse for 
$\hat M$ 
\bg\label{atonce}
\hat M^{-1}(z,z')=\tv e^{-i\hat l\lambda^2\phi(z)}G(z-z')
e^{i\hat l\lambda^2\phi(z')},\eg
where $G(z-z')$ is the periodic free Green's function defined
by\footnote{
This Green's function exists for $x_{||}\notin \Lambda$. }
\bg
\left(-i d_z-x_{||}\right)G(z-z')=\delta^2(z-z'),\eg
and has the Fourier series representation
\bg
G(z-z')=\tv^{-1}\sum_{\alpha\in\Lambda}
\frac{e^{i\alpha\cdot(z-z')}}{\alpha-x_{||}}.\eg
The inverse (\ref{atonce}) obviously satisfies 
$\hat M(z) \hat M^{-1}(z,z')=\tilde{\cal V}^2
\delta^2(z-z')$
for $z\neq \pm\omega$.
However, due to the singularities at $z=\pm \omega$ some caution 
is called for when
 interpreting
(\ref{atonce}) as the inverse of $\hat M$.
We will return to this point in the next section.
For now we will stick with (\ref{atonce}). 
$G(z)$ can be decomposed as follows
\bg
G(z)=\h\left(1-i\hat l\right)G_-(z)+\h\left(1+i\hat l
\right)G_+(z),\eg
where $G_\pm(z)$ are the following standard (i.e.
complex rather than quaternionic)
free
Green's functions
\bg
\left(-i\partial_{\mbox{z}}-\h\bar{\hbox{x}}_{||}\right)
G_+(z)=\h\delta^2(z),~~~
\left(-i\partial_{\bar{\mbox{z}}}-\h\mbox{x}_{||}\right)G_-(z)=\h
\delta^2(z).
\eg
Here
$\partial_{\mbox{z}}=\h(\partial_{z_0}-i\partial_{z_1})$,
$\mbox{x}_{||}=(x_{||})_0+i(x_{||})_1$
and the bar denotes complex conjugation.
Evidently
\bg
G_+(z)=\overline{G_-(-z)}.\eg
 
Now that we have the inverse of $\hat M$ (at least for $x_\perp=0$)
let us start the computation of the gauge potential
$A_\mu(x)$.
As was emphasized in the introduction it is not guaranteed that
$A_\mu(x)$ actually exists.
 We begin
 by considering $\rho(x)$ for
our putative one-instanton.
Inserting (\ref{atonce}) into (\ref{finaln}) yields
\begin{eqnarray}\label{n}
n(z)&=&\frac{\lambda\tv}{2}
\left[\left(1-i\hat l\right)e^{\lambda^2\left(
\phi(\omega)-\phi(z)\right)}G_-(\omega-z)\right.\\ \nonumber
&&\left.~~~~~~~~~+\left(1+i\hat l\right)
e^{-\lambda^2\left(
\phi(-\omega)-\phi(z)\right)}G_+(-\omega-z)\right].
\end{eqnarray}
We now appear to be in  trouble;
$\phi(z)\rightarrow\pm\infty$ as $z\rightarrow\pm\omega$,
and so $n(z)$ is proportional to the `infinite' constant
$e^{\lambda^2\phi(\omega)}$. Thus it appears that our use of the inverse 
(\ref{atonce}) 
was indeed unwarranted. 
Note that this problem is absent on $S^1\times\R^3$;
while the derivative of $\phi(z)$ is discontinuous at $z=\pm\omega$,
$\phi(\pm\omega)$ is well defined.
For now \sl we will proceed formally
 and
treat $\phi(\omega)=-\phi(-\omega)$ as if it were a finite constant.\rm\,
The integrand in
(\ref{integ})
is 
\begin{eqnarray}
\label{integrand}
n(z){n^\dagger(z)}&=&\frac{\lambda^2\tv^{2}e^{2\lambda^2\phi(\omega)}}{2}
\left[
\left(1-i\hat l\right)e^{-2\lambda^2\phi(z)}|G_-(\omega-z)|^2\right.\\
\nonumber
&&~~~~~~~~~~~~~~~~~~~~~~\left.
+\left(1+i\hat l\right)
e^{2\lambda^2\phi(z)}|G_+(-\omega-z)|^2
\right].\end{eqnarray}
Here $n^\dagger(z)=n^*(-z).$ 
Clearly the integrand (\ref{integrand})  has singularities
over and above the questionable
 $e^{2\lambda^2\phi(\omega)}$ factor.
We also note that  $n(z)n^\dagger(z)$
is not real.
Now we will argue that these singularities are \sl integrable \rm provided
\bg\label{scalelimit}
0<\lambda^2\tv<4\pi.\eg 
In the neighbourhood of $z=\omega$
we have the following singularity profile
\bg
|G_-(\omega-z)|^2
\propto\frac{1}{|\mbox{z}-\mbox{w}|^2},\quad
|G_+(-\omega-z)|^2 \quad\hbox{non-singular}.\eg
 $|G_-(\omega-z)|^2$ has a
 non-integrable singularity at $z=\omega$.
However, we must also consider the behaviour
of $\phi(z)$ at $z=\omega$
\bg
\phi(z)\sim -\frac{\tv}{4\pi}\log|\mbox{z}-\mbox{w}|.\eg
Near $z=\omega$ we have
\bg
|G_-(\omega-z)|^2e^{-2\lambda^2\phi(z)}\propto
|\mbox{z}-\mbox{w}|^{-2+\lambda^2\tv/(2\pi)}.\eg
This singularity is integrable for $\lambda^2>0$.
In fact if we take $\lambda^2\tv\geq 4\pi$ the singularity
disappears. However, then
$|G_-(\omega-z)|^2e^{-2\lambda^2\phi(z)}$ will not be integrable at
$z=-\omega$. Accordingly, for integrability
 at both $z=\omega$
and $z=-\omega$
we must impose (\ref{scalelimit}). 

 The bound  \eq{scalelimit} is nothing but the statement 
that $\lambda^2$,
the square of the ADHM size parameter,
 should not exceed the volume of the two-torus 
$\T^2$. 
Looking at the Abelian $U(1)$ potential $\hat A(z)$ the bound is quite
natural. Given that its associated field
strength is zero away from the fluxes one can formally
write it as a pure gauge,
i.e. $\hat A_i(z)=\partial_{z_i}\chi(z)$.
$\chi(z)$ is of course singular at the fluxes, but for
$0<\lambda^2\tilde{\cal V}<4\pi$ has a branch
cut joining the two fluxes.
At the critical value $\lambda^2\tilde{\cal V}=4\pi$
the branch cut disappears,
i.e. $\chi$ is single-valued on $\tilde\T^2$.
Then $\hat A(z)$ is truly a pure gauge and hence physically 
indistinguishable from the $\lambda^2\tilde{\cal V}=0$ case.

Let us now return to the problem of the infinite constant $e^{\lambda^2
\phi(\omega)}$ which seems to render our instanton meaningless.
Define a `finite' $n$ as follows
\bg\label{finite}
\lambda{\tilde{\cal V}}
n_f(z):=e^{-\lambda^2\phi(\omega)}n(z).\eg
For $x_\perp=0$ we have $n_f(z)=
\h\left(1-i\hat l\right)e^{-\lambda^2\phi(z)}
G_-(\omega-z)+
\h\left(1+i\hat l\right)e^{\lambda^2\phi(z)}
G_+(-\omega-z)$,
which is finite except at  the fluxes $z=\pm\omega$. 
The gauge potential can be written
\begin{equation}\label{finalpotential}
 A_\mu(x)=-
\frac{\int_{\tilde\T^2}
d^2z\left[
n_f(z)\partial_\mu
{n_f^\dagger(z)}-\partial_\mu\left(n_f(z)\right)
{n_f^\dagger(z)}
\right]}
{2\left(e^{-2\lambda^2\phi(\omega)}\lambda^{-2}
\tv^{-1}+\int_{\tilde \T^2}d^2 z
\,n_f(z){n_f^\dagger(z)}\right)},
\end{equation}
where the $\partial_\mu$ derivative is with respect to
$x_\mu$.
The only remnant of the infinite constant is the
$e^{-2\lambda^2\phi(\omega)}$ term in the denominator of
(\ref{finalpotential}); this exponential can be interpreted
as `zero', i.e. for our final potential we should take
\begin{equation}\label{finalpotential2}
A_\mu(x)=-
\frac{\int_{\tilde\T^2}
d^2z\left[
n_f(z)\partial_\mu
{n_f^\dagger(z)}-\partial_\mu\bigl(n_f(z)\bigr)
{n_f^\dagger(z)}
\right]}
{2\rho_f(x)},\end{equation}
where
\bg\rho_f(x)=\int_{\tilde \T^2} d^2z\,
n_f(z){n_f^\dagger(z)}.\eg
Although $n_f(z)n^\dagger_f(z)$ is not real a short calculation
suffices to express $\rho_f$ in a manifestly real and positive form
(here we use that $\phi(z)$ is an odd function, i.e. equation (\ref{odd}))
\bg
\rho_f(x_{||})=
\int_{\tilde\T^2}d^2z\,
e^{-2\lambda^2\phi(z)}|G_-(\omega-z)|^2.
\eg
So finally, the role of the infinite constant is simply
to expunge the $1$ from 
the definition of $\rho$. Without the $1$ the infinite
constant simply drops out of the final potential $A_\mu(x)$.
This is in sharp contrast to the situation on $S^1\times\R^3$,
where the 1 term \sl must \rm be kept since $\phi(\omega)$ is a finite
constant.

While (\ref{finalpotential2}) represents the final gauge potential
we have only given $n_f(z)$ and $\rho_f$
explicitly for the special
case $x_\perp=0$.
To construct $n_f(z)$ for $x_\perp\neq 0$ is non-trivial.
If we try to bring the $x_\perp$ inside the bracket of equation
 (\ref{rewrite})
we get
\bg
\tv^{-1}
\hat M(z)=e^{-i\hat l\lambda^2\phi(z)}
\left(-i d_z-x_{||}-x_\perp e^{-2i\hat l\lambda^2\phi(z)}\right)
e^{i\hat l\lambda^2 \phi(z)}.\eg
Proceeding as in the $x_\perp=0$ case we can write  the inverse
as follows
\bg\label{minv}
\hat M^{-1}(z,z')=\tv
e^{-i\hat l\lambda^2\phi(z)}\tilde G(z,z')e^{i\hat l\lambda^2\phi(z')},\eg
where
$\tilde G(z,z')$ is no longer a free Green's function
\bg\label{sings}
\left(-i d_z-x_{||}-x_\perp e^{-2i\hat l\lambda^2\phi(z)}\right)
\tilde G(z,z')=\delta^2(z-z').\eg 
Inserting (\ref{minv}) into (\ref{finite})  yields
\bg 
n_f(z)=\frac{1}{2}\left[
\left(1-i\hat l\right)\tilde G(\omega,z)
+\left(1+i\hat l\right)\tilde G(-\omega,z)\right]e^{i\hat l\lambda^2\phi(z)}
.\eg 
 A more detailed discussion of the properties of $n_f$ 
for $x_\perp\neq 0$ will be given elsewhere. 
 
The field strength derived from (\ref{finalpotential2}) is
\bg\label{fieldstrengthf}
 F_{\mu\nu}=\frac{\tv^{-1}}{\rho_f(x)}\int_{\tilde \T^2}d^2z
\int_{\tilde \T^2} d^2z'\,
n_f(z)\,\i_\mu\, f(z,z')\,\i_\nu^*\,
{n_f^\dagger(z')}-[\mu\leftrightarrow\nu],\eg
where $f(z,z')$ is
\begin{eqnarray}\label{greensfunctionff}
f(z,z')
&=&\tv^{-1}\int_{\tilde \T^2}d^2y
{\hat M}^{-1}(z,y)(\hat M^\dagger)^{-1}(y,z')\\ \nonumber
&&-\frac{\tv^{-1}}{\rho_f(x)}
\int_{\tilde \T^2}d^2y
{\hat M}^{-1}(z,y)n_f^\dagger(y)
\int_{\tilde \T^2}d^2y' n_f(y')(\hat M^\dagger)^{-1}(y',z').
\end{eqnarray}
Equations (\ref{fieldstrengthf}) and
(\ref{greensfunctionff}) are `finite'
forms of (\ref{fieldstrength}) and
(\ref{greensfunctionf}), respectively;
 as with the gauge potential
the $n(z)$ vector is replaced with its finite form, $n_f(z)$, and the
$1$ in $\rho$ is removed.

Since on the plane $x_\perp=0$ the explicit form of $n_f(z)$
and $\hat M^{-1}(z,z')$ are at hand we can also give a closed form for 
$f(z,z')$:
\begin{equation}\label{fformula}
f(z,z')=\h\left(1-i\hat l\right)f_-(z,z')+\h\left(1+i\hat l
\right)f_+(z,z'),
\end{equation}
where
\begin{equation}\label{c2}
f_\pm(z,z')=\tv
e^{\mp\lambda^2\phi(z)}
g_\pm(z,z')e^{\mp\lambda^2\phi(z')},
\end{equation}
and
\begin{eqnarray}\label{gdefinition}
g_\pm(z,z')&=&
\int_{\tilde\T^2}
 d^2y G_\pm(z-y)e^{\pm 2\lambda^2\phi(y)}G_\mp(y-z')\\ \nonumber
&&-\frac{1}{\rho_f}
\int_{\tilde \T^2}
 d^2y G_\pm(z-y)e^{\pm2\lambda^2\phi(y)}G_\mp(\pm\omega+y)
\\ \nonumber
&&~~~~~~~~~~~\times
\int_{
\tilde \T^2}
d^2y' G_\pm(\mp\omega-y')e^{\pm 2\lambda^2\phi(y')}G_\mp(y'-z').
\end{eqnarray}
A sufficient condition for the self-duality of $F_{\mu\nu}(x)$
is that $f(z,z')$ commutes with the quaternions. This condition is equivalent 
to  
\begin{equation}\label{thingtobeproved}
g_+(z,z')=e^{2\lambda^2\phi(z)}g_-(z,z')e^{2\lambda^2\phi(z')}.
\end{equation}
A (somewhat roundabout) proof of \eq{thingtobeproved} is given in 
Appendix~\ref{C}. 

To sum up, the gauge potential, $A_\mu(x)$, and hence 
the field strength, $F_{\mu\nu}(x)$,
can be written in terms of the `renormalised' $n_f(z)$.
We have explicitly determined $n_f(z)$ on the plane $x_\perp=0$.
At the point $x=0$ (i.e. $x_{||}=x_\perp=0$)
$n_f$ and hence $A_\mu$ is ill defined.
This is no surprise since we are working in the singular gauge
$u(x)=1$.
The singularity has its origins in the zero mode structure of the
$G_\pm(z)$; we can write
\begin{equation}\label{primegf}
G_+(z)=-\frac{1}{\tv \bar{\mbox{x}}_{||}}+
G_+'(z),~~~~~~~~~~~~~~
G_-(z)=-\frac{1}{\tv \mbox{x}_{||}}+
G_-'(z),\end{equation}
where the $G_\pm'(z)$
have no zero modes and are thus well defined for $x_{||}=0$.
Although $A_\mu$ diverges at $x=0$,  local
gauge invariants
such as $\mbox{tr} (F_{\mu\nu})^2$ (no sum)
should be smooth (presumably $C^\infty$).
As for the field strength itself, $F_{\mu\nu}(x)$,
this is not smooth at $x=0$,
but its components must be bounded.
Let us consider $F_{\mu\nu}$
at $x_\perp=0$ with $x_{||}\approx0$.
For $x_{||}\approx 0$ the zero modes in 
(\ref{primegf})
dominate and so we have\footnote{
Strictly speaking (\ref{nfapprox}) 
is only good away from $z=\pm\omega$. But as we are always dealing with 
integrable singularities we may safely employ (\ref{nfapprox})
under the integral sign.}
\begin{equation}\label{nfapprox}
n_f(z)\approx
-\frac{e^{-\lambda^2\phi(z)}}{2\mbox{x}_{||}\tv}\left(
1-i\hat l\right)-
\frac{e^{\lambda^2\phi(z)}}{2\bar{\mbox{x}}_{||}\tv}\left(1+i\hat l\right)
,\end{equation}
thus
\begin{equation}\label{rhofapprox}
\rho_f\approx
\frac{c}{|\mbox{x}_{||}|^2\tv^2},\end{equation}
where
\begin{equation}
c=\int_{\tilde\T^2} d^2z ~e^{2\lambda^2\phi(z)}.
\end{equation}
Plugging (\ref{nfapprox}) and (\ref{rhofapprox})
into the field strength formula (\ref{fieldstrengthf})
we see that in order to have a
bounded $F_{\mu\nu}$ in the vicinity of $x=0$,  
$f(z,z')$ must be well behaved for $x_{||}\approx 0$.
To see this consider, $F_{01}=F_{23}$, which for $x_\perp=0$
and $x_{||}\approx 0$ has the form
\begin{equation}\label{f01}
F_{01}\approx
-\frac{2\i_1\tv^{-1}}{c}\int_{\tilde \T^2}d^2z
\int_{\tilde \T^2}d^2z'~
e^{\lambda^2\phi(z)}
e^{\lambda^2\phi(z')}
f(z,z').\end{equation}
$F_{02}$ and $F_{03}$
are a bit more complicated;
here one finds 
phases of the form $\bar{\hbox{x}}_{||}/\mbox{x}_{||}$
which do not have definite values at $x_{||}=0$.
These phases are an artifact of the singular gauge;
$\mbox{tr}(F_{02})^2$ and $\mbox{tr}(F_{03})^2$ are
well behaved at $x_{||}=0$.
We now show that $f(z,z')$ is smooth in the vicinity
of $x_{||}\approx 0$.
Since the exponentials
in (\ref{c2}) are $x_{||}$-independent it suffices to show that
$g_+(z,z')$ has a well defined $x_{||}\rightarrow 0$ limit.
Glancing at (\ref{gdefinition}) one sees that the first term in $g_+(z,z')$
has double and single poles in $\mbox{x}_{||}$ and $\bar{\mbox{x}}_{||}$.
These poles are cancelled by the second term.
After some algebra one finds that
\begin{eqnarray}\label{gnear0}
g_+(z,z')&=&
\int_{\tilde \T^2}d^2y\left(
G_+'(z-y)-G_+'(-\omega-y)\right)e^{2\lambda^2\phi(y)}
\left(G_-'(y-z')-G_-'(y+\omega)\right)
 \nonumber\\
&&-\frac{1}{c}\int_{\tilde \T^2}d^2y~e^{2\lambda^2\phi(y)}
\left(
G_+'(z-y)-G_+'(-\omega-y)\right) \nonumber\\
&&~~~~~~~~~\times
\int_{\tilde\T^2}d^2y'~e^{2\lambda^2\phi(y')}
\left(G_-'(y'-z')-G_-'(y'+\omega)\right) \nonumber\\
&&+
{\cal O}(x_{||}),
\end{eqnarray}
which is well defined at $x_{||}=0$.
A similar expression can be obtained for $g_-(z,z')$.
From (\ref{c2})
the integrand in (\ref{f01}) is simply $g_+(z,z')$ and so all
we have to do is to integrate the right hand side of
(\ref{gnear0}) over $z$ and $z'$. Since the $G_\pm'(z)$ integrate
to zero this is  trivial. Putting all this together yields
\begin{eqnarray}
F_{01}&=&-\frac{2\i_1\tv^2}{c}\left[
\int_{\tilde\T^2}d^2y\,
e^{2\lambda^2\phi(y)}
|G_+'(-\omega-y)|^2\right.\\
\nonumber
&&~~~~~~~~~~~\left.-\frac{1}{c}\left|\int_{\tilde\T^2} d^2y\,
e^{2\lambda^2\phi(y)}G_+'(-\omega-y)\right|^2\right]
+{\cal O}(x_{||}).\end{eqnarray}
The content of the brackets
is strictly positive, i.e.\ we have not simply determined the field strength 
at a point where it is zero.

\subsection{Nahm transform interpretation}

In the previous section we implemented the ADHM construction
in the one-instanton sector for $\T^2\times\R^2$.
 However, in contrast to
 the caloron problem $n(z)$ appears not to exist.
 This was circumvented by formally extracting an infinite factor to obtain 
the `finite' $n_f(z)$.
Here we will explain precisely how the gauge potential (\ref{finalpotential2})
can be interpreted as the Nahm transform of the AB potential 
(\ref{abpotential}).
We would like to stress that this does \sl not \rm entail the kind of
formal manipulations we used to derive (\ref{finalpotential2})
in the first place
via the ADHM construction.

The Weyl operator on $\tilde\T^2$ associated with $\hat A(z)$ has
 \sl two \rm square integrable zero modes
\footnote{In ref \cite{periodicmonopoles}
where the dual torus  was take to be
$\tilde S^1\times\R$ a limiting
case of $\tilde \T^2$, $\hbox{dim}(
\hbox{ker}\hat D^\dagger)=2$
was also obtained.}. These modes 
can be identified with the columns of $n_f^\dagger(z)$ when the quaternionic
object $n_f(z)$ is recast as a $2\times 2$ matrix with complex entries.
To set the scene let us briefly recall how the Nahm transformation
is formulated on $\T^4$. Consider a self-dual $SU(N)$ potential
$A_\mu(x)$ on $\T^4$ with instanton number $k$.
Then one studies the Weyl operator associated with the $U(N)$ potential
obtained by adding a constant abelian potential
$- i z_\mu$ to $A_\mu$
\begin{equation}
D_z(A)=i_\mu D_z^\mu(A), ~~~~~~
D_z^\mu=\partial^\mu+A^\mu(x)- iz^\mu.
\end{equation}
Provided certain mathematical technicalities are met
$D^\dagger=-i_\mu^* D_z^\mu(A)$
has $k$ square integrable zero modes
$\psi_z^i(x)$ with $i=1,2,...,k$. For convenience we 
take them to be normalised to unity.
The $U(k)$ potential
\begin{equation}\label{nahmpotential}
\hat A^{ij}_\mu(z)=\int_{\T^4}
d^4x \,{\psi_z^i}^\dagger(x)
\frac{\partial}{\partial z^\mu}\psi^j_z(x),
\end{equation}
is a self-dual potential on the dual torus
$\tilde \T^4$ with instanton number $N$.
On $\T^4$ this procedure is involutive and (in a suitable gauge)  free of 
singularities.

Let us  write
the Weyl operator associated with the AB potential  
(\ref{abpotential}) as a $2\times 2$ matrix:
\begin{equation}
-\frac{i}{2}D_x^\dagger(\hat A)=
S\btensor{(}{cc}
i\partial_{\bar{\hbox{z}}}+\h\hbox{x}_{||}-i\partial_{\bar{\hbox{z}}}\phi&
\h\hbox{x}_\perp\di \\
-\h\bar{\hbox{x}}_\perp&
i\partial_{\hbox{z}}+\h\bar{\hbox{x}}_{||}+i\partial_{\hbox{z}}\phi
\etensor{)}S^{-1},
\end{equation}
where 
\footnote{
$S$ is a unitary transformation with the property
$S^{-1}\sigma_1 S=\sigma_3$,
$S^{-1}\sigma_2 S=\sigma_2$ and $S^{-1}\sigma_3S=-\sigma_1$.}
$S=(\id-i\sigma_2)/\sqrt{2}$ and
$\hbox{x}_\perp=x_2+ix_3$.
For $\hbox{x}_\perp=0$ one can write down \sl two \rm 
square-integrable
zero modes
for $D_x^\dagger(\hat A)$
\begin{equation}
\psi_x^1(z)=
\frac{1}{\sqrt {\rho_f}}S\btensor{(}{c}
e^{\lambda^2\phi(z)}G_-(z+\omega) \\\di  
0\etensor{)},~~~~~~
\psi_x^2(z)=
\frac{1}{\sqrt {\rho_f}}S\btensor{(}{c}
0 \\\di  e^{-\lambda^2\phi(z)}G_+(z-\omega)\etensor{)}.
\end{equation}
Both zero modes are singular at $z=\pm\omega$.
Inserting these (normalised) zero modes into (\ref{nahmpotential})
yields exactly the same potential 
(discarding the $U(1)$ part of the U(2) connection)
as constructed in the previous section.
If one writes $n_f^\dagger$ as a $2\times 2$ matrix the columns
are (upto a normalisation factor) the Nahm zero modes.
As should be clear from the considerations of the previous section
it is non-trivial to obtain the zero modes for
 $\hbox{x}_\perp\neq 0$.
The crucial feature of these zero 
modes is that although they are singular at the fluxes
 $z=\pm\omega$ the Weyl equation does \sl not \rm have sources,
i.e. $D_x^\dagger(\hat A)\psi_x^i(z)$ is \sl exactly \rm zero.
Basically, the damping exponentials soften the singularities of the Green's
 functions
$G_-(z+\omega)$
and $G_+(z-\omega)$ so that no delta function sources occur on the right
 hand side of the Weyl equation.

It is also instructive to compare the situation on $\T^2\times\R^2$ 
with the caloron case ($S^1\times \R^3$).  
It is easy to write down the corresponding zero modes on $\tilde S^1$
for the caloron problem. One simply replaces the $\tilde\T^2$ Green's 
functions $\phi$, $G_+$ and $G_-$ with their $\tilde S^1$ counterparts.
However, in this case the Weyl equations \sl do \rm have sources.
The $e^{\pm\lambda^2\phi(z)}$, being finite at $z=\pm\omega$, 
have no damping effect on the $G_\pm$.
Because of these sources, direct
insertion of the $\tilde S^1$ `zero modes'
into (\ref{nahmpotential}) does \sl not \rm yield a self-dual potential on
$S^1\times \R^3$. Rather, one has to change the \sl normalisation 
\rm of the zero modes to compensate for the sources.
This amounts to including $1$ in the definition of $\rho$.

Given that the $\tilde \T^2$ Weyl operator has perfect zero modes what
 exactly is the status of the inverse of
$\hat M$ introduced in the previous section?
What is clear is that our $\hat M^{-1}(z,z')$ is \sl not \rm the inverse of 
$\hat M$ on the space of square integrable spinors;
no such inverse exists.
Our  $\hat M^{-1}(z,z')$ can be viewed as the inverse of $\hat M$
 on a space of functions on $\tilde \T^2$ having
softer singularities at the fluxes than the zero modes.
In any case $\hat M^{-1}(z,z')$ only enters at intermediate stages of the 
calculation.
What is important is $n_f(z)$, which, as we have shown here, encodes
 two perfect zero modes of our Weyl operator.

Thus it seems there are three types of Nahm transformation.
First and foremost is the $\T^4$ transformation where all potentials and
attendant zero modes are smooth.
For $\T^n\times \R^{4-n},~~~n<4$ the self-duality equations on $\tilde\T^n$
have source terms.
The Weyl zero modes on $\tilde \T^n$ are also singular but for $n=2$
(and presumably $n=3$) there are no source terms in the Weyl equation
and so (\ref{nahmpotential}) can be applied without modification.
For $n=1$ (and $n=0$ for that matter) the Weyl equation has source terms 
which are finessed by
altering the normalisation of the zero modes.

\setcounter{equation}{0}
\section{Two-instantons}\label{sec4}

The two-instanton problem on the torus presents new challenges.
In particular, the Nahm potential, $\hat A(z)$, on $\tilde\T^n$
is non-Abelian; for $k=2$ instantons $\hat A(z)$
is an $SU(2)$ potential.
In contrast to the one-instanton
 case the determination of
$\hat A(z)$ is itself a non-trivial exercise.
For $\T^2\times\R^2$ and $S^1\times\R^3$
the field strength associated with the Nahm potentials is
zero, except at the singularities.
But even here we do not have closed forms 
for $\hat A(z)$.
In section 4.1 we give some \sl particular \rm solutions
to the $k=2$ ADHM constraints.
The associated Weyl equations for the $\T^2\times\R^2$ problem are 
investigated
in section 4.2. This analysis is very similar to that of section
3.2 for the one instantons.
Indeed, the resulting two-instantons can be viewed as \sl twisted
\rm one instantons when the torus is  cut in half.

\subsection{ADHM constraints on $\T^n\times \R^{4-n}$}

In the previous chapter we considered the general one-instanton
which (apart for $S^1\times\R^3$) is  non-periodic.
For $k=2$ the ADHM constraint (\ref{cancontor})
is obviously more complicated.
In particular, the quadratic term in (\ref{cancontor})
is, in general, non-zero.
There is however one simplification at the two-instanton level;
there exist non trivial
solutions of the ADHM constraints which correspond
to periodic gauge potentials on $\T^n\times\R^{4-n}$.
This is because we can choose the two `component' instantons
to have a different orientation in group space.

\begin{figure}[ht]
\begin{minipage}[ht]{16cm}
\centerline{\epsfysize=6 cm\epsffile{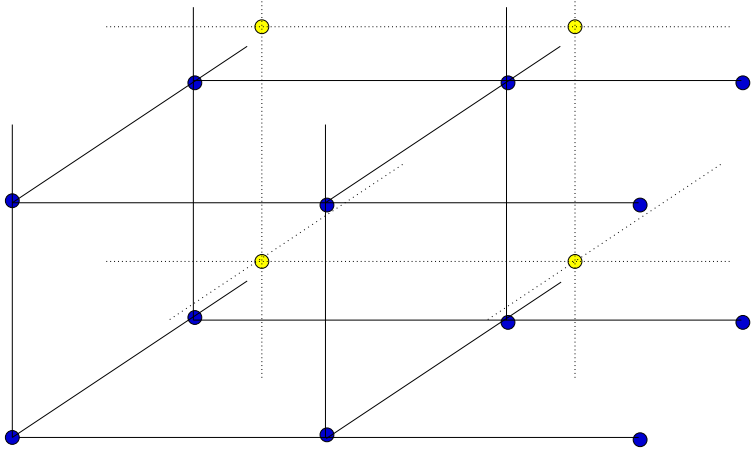}}
\caption{\label{NaCl}\textsl{
One `species' of instantons lying at the midpoints of the lattice points 
of the other species of instantons.}}
\end{minipage} \end{figure}

For simplicity, let us restrict ourselves to the periodic case.
Then for $k=2$ we can write $v$ and $\hat M$ as follows
\bg
v=(v_1^\alpha\,\, v_2^\alpha),\quad
\hM=\btensor{(}{cc}
\hM_{11}^{\alpha \beta}&\hM_{12}^{\alpha \beta}\di \\
\hM_{21}^{\alpha \beta}&\hM_{22}^{\alpha \beta}
\etensor{)},
\eg
where $v_1^\alpha=q_1$, $v_2^\alpha=q_2$, and
\bm{ll}
\hM^{\alpha \beta}_{11}=\delta^{\alpha \beta}(\alpha+y_1-x)
+b_{11}^{\alpha \beta}, &\di  \quad
\hM^{\alpha \beta}_{12}=\hM^{\beta \alpha}_{21}=b^{\alpha \beta}_{12}\\\di 
\hM^{\alpha \beta}_{22}=\delta^{\alpha \beta}(\alpha+y_2-x)
+b_{22}^{\alpha \beta}. &\di 
\eg
We now have to determine the $b$ matrices via (\ref{cancontor}).
In the one instanton calculation we relied on the vanishing of the
quadratic term in (\ref{cancontor}).
While this will not hold, in general, for the two instanton case
there may be \sl particular \rm solutions where the quadratic term is
zero.
Indeed on $\R^4$, the $k=2$ problem is expedited by the vanishing
of the quadratic term in (\ref{cancon}) \cite{cws}.
If the quadratic term in (\ref{cancontor}) is zero,  the
$b$ matrices read
\bg\label{exact}
b_{11}^{\alpha \beta}=b_{22}^{\alpha \beta}=0,
\quad
b_{12}^{\alpha \beta}=-\frac{1}{2(\alpha-\beta+y_1-y_2)^*}Q,\eg
where
\bg
Q=q_1^*q_2-q_2^*q_1.\eg
In Appendix A we will prove that  if $2(y_1-y_2)\in\Lambda$ 
and $y_1-y_2\notin\Lambda$
then the quadratic term does indeed vanish.
For example this happens for $y_1-y_2=\h (e_0+e_1+...+e_{n-1})$.
This means that the lattice points of the second `species' of
instanton lie
exactly at the midpoints (see figure \ref{NaCl}) 
of the lattice points of the first. 

In the special case $n=1$ (i.e. the caloron problem)
one only needs $y_1-y_2$ to be parallel to $e_0$
for the quadratic term to vanish. This is a consequence of the fact that for
$S^1\times\R^3$ one may take $e_0$ and hence the elements of $\Lambda$ to
be real.
For $n>1$, $2(y_1-y_2)\in\Lambda$ is a \sl necessary \rm\, condition
for the vanishing of the quadratic term. Thus for $2(y_1-y_2)\notin
\Lambda$ (\ref{exact}) is an approximation;
(\ref{exact}) is then the first term of a power series 
expansion in the scale parameters.

Let us concentrate on the cases where the quadratic terms does vanish.
Fourier transformation yields
 $\tv^{-1}\hat M=-id_z-x+\hat A(z)$, where
$\hat A(z)$ is the  $SU(2)$ potential 
\bg\label{ahat}
-i\hat A(z)=\btensor{(}{cc}
\h(y_1-y_2)& 
\h %in my calculations we need a factor one half at this point
ie^{-i(y_1-y_2)\cdot z} d_z\psi(z)Q\\
-
\h%in my calculations we need a factor one half at this point
ie^{i(y_1-y_2)\cdot z} d_z\psi(-z)Q&\h(y_2-y_1)\etensor{)},
\eg
and 
\bg 
\psi(z)=\sum_{\alpha\in\Lambda}
\frac
{e^{i(\alpha+y_1-y_2)\cdot z}}{|\alpha+y_1-y_2|^2}.
\eg
 $\psi(z)$ is a Green's function for the Laplace
operator on $\tilde \T^{n}$
\bg
 d_z d_z^*\psi(z)=-\tv e^{i(y_1-y_2)\cdot z}
\delta^{n}(z).\eg
Observe that $\psi$ is non-periodic
\bg
\psi(z+\tilde e_i)=e^{i (y_1-y_2)\cdot\tilde e_i}\psi(z),\eg
where $\tilde e_i$ refers to the dual basis;
$\tilde e_i\cdot e_j=2\pi\delta_{ij}$.
Now if $2(y_1-y_2)\in\Lambda$ and $(y_1-y_2)\notin\Lambda$, 
$\psi(z)$ will be antiperiodic
in at least one direction, and periodic in the remaining
 directions.
One can also see that for these special values of $y_1-y_2$,
$\psi(z)$ is real. The reality of $\psi$ is a sufficient condition for
the potential (\ref{ahat}) to be self-dual.

We now appear to have to deal with a non-Abelian Weyl operator. 
In what follows the inversion problem is reduced 
to an Abelian problem much like that for the one instanton case.
Of course, in the light of the previous chapter 
due care
 regarding the 
meaning of the inverse is in order.  
$ \hM$ can be rewritten as follows
\bg\label{projection}
\tv^{-1}\hM=e^{-\frac{i}{2}(y_1-y_2)\cdot z\sigma_3}
P^{-1}\btensor{(}{cc}D^+&0\cr 0&D^-\etensor{)}P
e^{\frac{i}{2}(y_1-y_2)\cdot z \sigma_3},\eg
where $D^\pm$ are the (Abelian) Weyl operators
\bm{ccc}\label{abweyl} 
D^\pm=
-i d_z-x\pm 
\h 
 d_z \psi\, Q, & &\di P=\frac{1}{\sqrt 2}(\id+i\sigma_1).\eg
 The inverse of $\hat M$ is simply
\bg
\hat M^{-1}(z,z')=
\tv
e^{-%replaced + by -
\frac{i}{2}(y_1-y_2)\cdot z\,\sigma_3}
P^{-1}
\Delta(z,z')Pe^{%replaced - by +
\frac{i}{2}(y_1-y_2)\cdot z'\sigma_3},\eg
where $\Delta(z,z')$ is a Green's function for the diagonal
operator $\hbox{diag}(D^+,D^-)$.
Note that the exponentials in the decomposition of $\hat M^{-1}
(z,z')$ are not periodic.
To ensure a periodic $\hat M^{-1}(z,z')$
we must impose certain non-periodic boundary conditions
on $\Delta(z,z')$.
Since we require $\hat M(z) \hat M^{-1}(z,z')=\tv^2
\delta^n(z-z')$,  then it follows that
\bg\label{source}
\btensor{(}{cc}D^+&0\cr
0&D^-\etensor{)}\Delta(z,z')=Pe^{\frac{i}{2} (z-z')\cdot(y_1-y_2)\sigma_3}
P^{-1}\delta^{n}(z-z')
.\eg
It is convenient to absorb the exponential factor into the delta
function.
That is, consider the following (non-periodic) delta functions
\bg
\delta_1^n(z)=e^{%replaced - by +
\frac{i}{2} z\cdot(y_1-y_2)}\delta^n(z),\quad
\delta_2^n(z)=e^{-%replaced + by - 
\frac{i}{2} z\cdot(y_1-y_2)}\delta^n(z).\eg
Using the following four (Abelian) Green's functions, 
$\Delta^\pm_i(z,z'), \quad i=1,2$,
where
\bg
\label{greens}
D_z^\pm \Delta^\pm_i(z,z')=\delta_i^n(z-z').\eg
$\Delta$ can be written as 
\bg
\label{invdiag}
\Delta(z,z')= \h%here a factor 1/2 inserted
\btensor{(}{cc}
\Delta^+_1+\Delta^+_2
&
i\left(\Delta^+_1-\Delta^+_2\right)\cr
-i\left(\Delta^-_1-\Delta^-_2\right)&
\Delta^-_1+\Delta^-_2
\etensor{)}(z,z').\eg
Accordingly
\bg
\label{fullinverse}
\hat M^{-1}(z,z')=
\frac{\tv}{2} e^{-\frac{i}{2}z\cdot(y_1-y_2)\sigma_3}\btensor{(}{cc}
\Delta^+_1+\Delta^-_1&
-i\left(\Delta^-_2-\Delta^+_2\right)\cr
i\left(\Delta^-_1-\Delta^+_1\right)&
\Delta^+_2+\Delta^-_2
\etensor{)}(z,z')
e^{\frac{i}{2} z'\cdot(y_1-y_2)\sigma_3}.\\ \eg

\subsection{Two-instanton on $\T^2\times\R^2$}

Much as in section 3.2
 we may take $e_0$ to be real and
$e_1$
to be proportional to $Q$.
Thus $\hat Q=Q/|Q|$ plays  the same role as $\hat l$
did in the previous section. Indeed,
the analogue of (\ref{complexcoords}) is just
$z=\h\left(1-i\hat Q\right)\mbox{z}+\h\left(1+i\hat Q
\right)\bar{\mbox{z}}$.
We can write the Abelian Dirac operators $D^\pm$ defined in (\ref{abweyl})
as follows
\bg\label{abweyl2}
D^\pm=e^{\mp \h i Q\psi(z)}\left(
-i d_z-x_{||}\right)e^{\pm \h iQ\psi(z)}-x_\perp.\eg
 For the case $y_2-y_1=\h (e_0+e_1)$, we have
\def\lo2{\frac{L_0}{4\pi}}
\bg
\psi(z)=\frac{\tv}{4\pi}
\log\frac{\left|\theta\left(
\lo2  \mbox{z}+\frac{iL_0}{4L_1},\frac{iL_0}{2L_1} \right)
\right|^2}
{\left|\theta\left(\lo2 \mbox{z}+\h,\frac{iL_0}{2L_1}
\right)\right|^2
} ,
\eg
which is antiperiodic in both directions.

When $x_\perp=0$, the four  Green's functions 
$\Delta_i^\pm$ read\footnote{Note that $\Delta^\pm_i(z,z')=
e^{\mp i Q\psi(z)}G_i(z-z')e^{\pm i Q\psi(z')}$
is not correct, since one has to take into account the
non-periodicity
of the exponentials $e^{\pm i Q\psi}=\cosh\left(|Q|\psi\right)\pm
i\hat Q\sinh\left(|Q|\psi\right)$.}
\begin{eqnarray}
\label{massless}
\Delta^\pm_1(z,z')&\!\!=&e^{\mp \h i Q\psi(z)}\left[G_1(z-z')
\cosh\left(\h |Q|\psi(z')\right)
\pm G_2(z-z') i\hat Q\sinh\left(\h |Q|\psi(z')\right)\right]\nonumber \\ 
\Delta^\pm_2(z,z')&\!\!=& e^{\mp \h i Q\psi(z)}\left[G_2(z-z')
\cosh\left(\h |Q|\psi(z')\right)
\pm G_1(z-z') i\hat Q\sinh\left( \h |Q|\psi(z')\right)\right]
,\nonumber \\
&&\end{eqnarray}
where the $G_i(z-z')$
are (non-periodic) \sl free \rm Green's functions defined as
\bg
\left(
-i d_z-x_{||}\right)G_i(z-z')=\delta_i^2(z-z'),\quad i=1,2.\eg
Inserting (\ref{massless}) into (\ref{fullinverse})
 yields
\bg
\hat M^{-1} (z,z')=\tv \Psi(z)
\btensor{(}{cc}
G_1(z-z')&0\cr
0&G_2(z-z')\etensor{)}
\Psi^{-1}(z'),
\eg
where $\Psi(z)$ is the $2\times 2$ matrix
\bg
\Psi(z)=\btensor{(}
{cc}
e^{- \h i (y_1-y_2)\cdot z}\cosh\left(\h |Q|\psi(z)\right)&
\hat Q e^{- \h i (y_1-y_2)\cdot z}\sinh\left(\h |Q|\psi(z)\right)
\cr
-\hat Q e^{\h i (y_1-y_2)\cdot z}\sinh\left(\h |Q|\psi(z)\right)&
e^{\h i (y_1-y_2)\cdot z}\cosh\left(\h |Q|\psi(z)\right)
\etensor{)}.\eg
The two component row vector $n(z)$ is 
\bg
n(z)=\tv(q_1\, , \, q_2)
\Psi(0)\btensor{(}{cc}
G_1(-z)&0\cr
0&G_2(-z)\etensor{)}\Psi^{-1}(z).\eg
Again we encounter infinite constants; 
$\psi(z)\rightarrow\infty$ as 
$z\rightarrow 0$ and so
all entries of the matrix $\Psi(0)$
are `infinite'.
As in section~\ref{6} we will temporarily treat $\Psi(0)$ as a finite
object.
In the light of our one instanton calculation we expect some
constraints on $q_1$ and $q_2$.
We can choose $q_1$ to be real.
In appendix B we show that for $n(z){n^\dagger(z)}$
to be integrable requires that 
\bg\label{2con1}
(q_1,\,\,\, q_2)=\lambda(1,\,\,\, \hat Q),\eg
where $\lambda$ is a \sl common \rm scale parameter since
$|q_1|=|q_2|=\lambda$. Observe that \sl the relative group orientation
of the two instantons is fixed\rm. 
If the orientation of the first instanton
lies at the `North pole' of $S^3\equiv SU(2)$, then the orientation of
the second instanton sits on the equator.
 Much as in the one instanton case the absence of non-integrable
singularities 
leads to an upper bound on the scale parameter
\bg\label{2con2}
0<\lambda^2\tV<2\pi.\eg 
Another consequence of (\ref{2con1})
is that $(q_1,\,\, q_2)$ is an eigenvector of the infinite matrix
$\Psi(0)$, i.e. $(q_1,\,\, q_2)\Psi(0)=e^{\h|Q|\psi(0)}(q_1,\,\,
q_2)$. As in the one instanton calculation we define a `finite'
row vector $\lambda\tv
n_f(z)=e^{-\h|Q|\psi(0)}n(z)$.
The final gauge potential is obtained by
replacing $n(z)$ with $n_f(z)$ in \eq{amu} and replacing \eq{rho1} 
with $\rho=\tv^{-1}\rho_f=\tv^{-1}\int_{\tilde \T^2} n_f(z){n^\dagger_f
(z)}.$

In the course of the construction a number of constraints
have been put on the ADHM data.
 It is helpful to divide these constraints into two.
The first constraints are simply those imposed by hand to achieve
technical simplification,
i.e. we imposed periodicity and the midpoint condition in order that
we could exactly determine the Weyl operator.
In addition to these constraints we were \sl forced \rm to impose the
additional constraints 
\eq{2con1} and (\ref{2con2}).
By virtue of the midpoint prescription
and (\ref{2con1}) our two instantons begin to resemble one instantons
if we cut $\T^2$ in half.
In fact if we had chosen $y_1-y_2=\h e_0$ or $y_1-y_2={1\over 2}e_1$
instead of $y_1-y_2=\h(e_0+e_1)$ then our `two instanton'
would be nothing more than a `doubled' one instanton. That is one
can always produce a two-instanton on $\T^n\times\R^{4-n}$
by taking a one instanton and doubling one of the periods.
To show this equivalence one simply compares the `two instanton'
with $y_1-y_2=\h e_0$ or $y_1-y_2=\h e_1$
with the one instanton with $\omega={1\over 4} \tilde e_0$ or
$\omega={1\over 4} \tilde e_1$. Then using the $q_i\rightarrow -q_i$ symmetry
mentioned at the end of section~\ref{fourier} one can show that the two sets
of ADHM data correspond to the same instanton.
The two instanton corresponding to $y_1-y_2=\h(e_0+e_1)$
appears to be `genuine' in the sense it is not equivalent to
some one-instanton solution.
However it seems plausible that the $y_1-y_2=\h(e_0+e_1)$
case corresponds to a twisted one instanton (the twisted Nahm transformation
is discussed in  \cite{arroyo}).

\setcounter{equation}{0}
\section{Discussion}\label{sec5}

In this paper we have described in a general way
how to implement the ADHM construction of $SU(2)$ instantons on
$\T^n\times\R^{4-n}$.
The first step (which corresponds to solving the quadratic
ADHM constraint)
is to construct a self-dual $SU(k)$ ($U(1)$ for $k=1$)
potential, $\hat A(z)$,
 on the dual torus $\tilde\T^n$ (here $k$ is the
topological charge).
$\hat A(z)$ has singularities which are determined by
the ADHM data (i.e. the scales, positions and group orientation
of the `component' instantons).
We have constructed the Weyl operators corresponding to the general
one-instanton and some two instantons on $\T^n\times\R^{4-n}$.
However, the problem of solving the  Weyl equations
poses a considerable technical challenge.
One is therefore motivated to start with lower values of
$n$. 
We have considered the $n=2$ problem in some detail.

The solutions here are not deformations of 't Hooft
instantons;
the 't Hooft ansatz fails to provide solutions on $\T^2\times\R^2$.
Unlike for $S^1\times\R^3$ we are forced to impose constraints on the
ADHM parameters in order to guarantee a well defined potential on
$\T^2\times\R^2$.
In particular, we find an upper bound 
on the scale parameters;
for the one-instanton,
$\lambda^2\tilde{\cal V}<4\pi$
and for our restricted two-instanton we found
that $\lambda^2\tilde{\cal V}<2\pi$ (here we were forced to 
give the two component instantons a common scale parameter).
%One is tempted to speculate that in general the $\T^2\times\R^2$ 
%instantons satisfy
%$k\lambda^2\tilde{\cal V}<4\pi$.

%For the 2-instanton  we restricted ourselves to
%the periodic case, and then imposed the `midpoint' constraint
%$2(y_1-y_2)\in\Lambda$. It would be interesting to relax these 
%constraints. However, we would like to point out 
%that to obtain the bound we only needed to study $n_f$ in the neighbourhood 
%of the poles. This suggests that this argument holds true 
%in the general case. 
%We expect the general $k=3$ (and higher charge) calculation
%to be prohibitively difficult, although particular solutions
%may be within reach.

For $n>2$, i.e. $\T^3\times\R$ and $\T^4$,
 the Weyl equations seem more problematic.
While the $\T^2\times\R^2$ Weyl operator corresponds to an
Aharonov-Bohm problem on $\tilde\T^2$,
on $\T^3\times\R$ we have to solve the Weyl equation on
$\tilde \T^3$ in the (self-dual) 
background of an electric and magnetic dipole
field
\cite{vanbaalletter}.
 For $ \T^4$ the one instanton calculation should fail.
Presumably there is no way to avoid non-integrable singularities.
For our restricted two instantons the prospects seem a little brighter.
This is because these seemingly
correspond to twisted one instantons (or even $\h$
instantons in the presence of non-orthogonal twists).
There is no known obstacle to the existence of such 
objects on $\T^4$.

Although the $\T^3\times \R$ and $\T^4$ problems certainly merit
more attention the $\T^2\times\R^2$ case
requires further development.
Even in the 1-instanton sector we were only able to provide closed forms
 for $A_\mu(x)$
and $F_{\mu\nu}(x)$ in a 2-dimensional subspace ($x_\perp=0$)
of $\T^2\times\R^2$. To obtain analytic results for $x_\perp\neq0$
requires progress in dealing with  massive Aharonov-Bohm type
Dirac equations
on $\tilde\T^2$.
Furthermore, we have said nothing about the geometry of the moduli
space or the constituent monopoles of our instantons.
%That we see a limit on the scale parameter indicates that the
%1-instanton moduli space is the base manifold times a compact space.
One could numerically plot the action density
of the one instantons in the plane $x_\perp=0$
to see if there  are two peaks
associated with the two expected monopole constituents.

\section*{Acknowledgements}

C. F. is grateful to C. J. Biebl for helpful discussions.
We thank P. van Baal for his  comments
on a preliminary version of the manuscript.
Part of the research of T. T. was performed during his
stay
at the Institute of Theoretical Physics in Jena,
and in the latter stages of the work he was supported by the
Deutsche Forschungsgemeinschaft (grant DFG-Re 856/4-1).
In the early stages of this work C. F. was supported by the
DFG
(grant DFG-Wi 777/3-2).

\vskip 10pt

\appendix
\setcounter{equation}{0}
\section{The quadratic term in (\ref{cancontor})}\label{A}

In this appendix we show that the quadratic term in (\ref{cancontor})
vanishes for the one instanton and particular two instanton 
described in chapter~\ref{sec4}.
 
Let us start with the one instanton.
The quadratic term in question is
\bg\label{quadone}
{\cal R}^{\alpha\beta}=\sum_{\gamma\in\Lambda}
\left(
b^{\gamma\alpha}{}^* b^{\gamma\beta}-b^{\gamma\beta}{}^*
b^{\gamma\alpha}
\right).\eg
Assuming ${\cal R}^{\alpha\beta}=0$ leads to (\ref{bmatrix}).
Inserting this into 
(\ref{quadone})
gives
\bg\label{explicitquadone}
{\cal R}^{\alpha\beta}& =&\di \!\!-\lambda^4\sum_{\gamma\in
  \Lambda\setminus
\{\alpha,\beta\}}\hat l
\left(
\frac{1}{(\gamma-\alpha)^*}\frac{1}{\gamma-\beta}
-\frac{1}{(\gamma-\beta)^*}\frac{1}{\gamma-\alpha}
\right)\hat l\\\di 
& &\di \ \ \ \ \  
~~~~~~~~~~~~~\times\sin\left[(\alpha-\gamma)\cdot\omega\right]
\sin\left[(\beta-\gamma)\cdot\omega\right].\eg
It is clear that each summand in 
(\ref{explicitquadone}) does not separately vanish.
Rather there is a pairwise cancellation;
for each $\gamma\in \Lambda\setminus\{\alpha,\beta\}$
there is exactly one other lattice point $\gamma'\in
\Lambda\setminus\{\alpha,\beta\}$
so that the two summands add up to zero.
It is apparent that the appropriate choice for $\gamma'$
is
$\gamma'=-\gamma+\alpha+\beta.$
If $2 \gamma = \alpha+\beta$, i.e. $ \gamma' = \gamma $, then 
the summand itself vanishes.

The argument is similar for the two instanton of section~\ref{sec4}.
Here  the quadratic term is 
\bg
{\cal R}_{ij}^{\alpha\beta}=
\sum_{\gamma\in\Lambda}
\left(
b_{1i}^{\gamma\alpha}{}^*
b_{1j}^{\gamma\beta}-
b_{1j}^{\gamma\beta}{}^*
b_{1i}^{\gamma\alpha}
+
b_{2i}^{\gamma\alpha}{}^*
b_{2j}^{\gamma\beta}-
b_{2j}^{\gamma\beta}{}^*
b_{2i}^{\gamma\alpha}
\right).\eg
Inserting (\ref{exact}) gives
${\cal R}_{12}^{\alpha\beta}={\cal R}_{21}^{\alpha\beta}=0$,
and
\begin{eqnarray}\label{quadratic}
{\cal R}_{22}^{\alpha\beta}&=&
\sum_{\gamma\in\Lambda}
\left(
b_{12}^{\gamma\alpha}{}^*
b_{12}^{\gamma\beta}-
b_{12}^{\gamma\beta}{}^*
b_{12}^{\gamma\alpha}
\right)\\ \nonumber
&=&-\frac{1}{4}\sum_{\gamma\in\Lambda}\left(
Q\frac{1}{
\gamma-\alpha+y_1-y_2}
\frac{1}{(\gamma-\beta+y_1-y_2)^*}Q\right.
\\ \nonumber
&&\left.~~~~~~~~~~~-Q\frac{1}{
\gamma-\beta+y_1-y_2}
\frac{1}{(\gamma-\alpha+y_1-y_2)^*}Q
\right).
\end{eqnarray}
Now we will show that ${\cal R}_{22}$ is zero for
$2(y_1-y_2)\in\Lambda$. As in the one instanton case
 each summand in (\ref{quadratic})
does not separately vanish. For 
 each $\gamma\in\Lambda$ there is  one  other
lattice point $\gamma'\in\Lambda$ so that the two summands add up
to zero
\bg
\gamma'=-\gamma+\alpha+\beta-2(y_1-y_2).\eg
Since $\gamma'\in\Lambda$ we require $2(y_1-y_2)\in\Lambda$.
If $2\gamma=\beta+\alpha-2(y_1-y_2)$ then $\gamma'=\gamma$
so that we do not have two counterbalancing summands. However, in 
this case the summand itself vanishes.

\setcounter{equation}{0}
\section{ Equation (\ref{thingtobeproved})}\label{C}

In this appendix we outline a proof of (\ref{thingtobeproved})
which, for $x_\perp=0$, is equivalent to the statement that
$f(z,z')$ commutes with the quaternions.
In the caloron problem one simply notes that $f$ is the inverse
of $M^\dagger M$ which by construction commutes with the quaternions.
We could also explicitly
check that our $f$ is the inverse of $M^\dagger M$.
However, we would face the thorny problem of coincident fluxes and sources
\cite{sources,sources2,ds}.
 Therefore, we will adopt a more pedestrian approach.
Before we embark on this we note that for
$z+z'=0$ a trivial change of variables in the integrals
defining $g_-(z,z')$ suffices to verify (\ref{thingtobeproved}).
For $z+z'\neq 0$ we have a more indirect argument.
When $z\neq \omega$ it is easy to check that
\begin{equation}
\left(-i\partial_{\bar{\hbox{z}}}-\h\hbox{x}_{||}\right)
e^{-2\lambda^2\phi(z)}
\left(-i\partial_{\hbox{z}}-\h \bar{\hbox{x}}_{||}\right)
\left(g_+(z,z')-e^{2\lambda^2\phi(z)}
g_-(z,z')e^{2\lambda^2\phi(z')}\right)=0.
\end{equation}
This shows that the left and right hand sides of (\ref{thingtobeproved})
satisfy the same differential equations.
To complete the argument we must show that they obey the same boundary
 conditions.
Clearly both are periodic on $\tilde\T^2$,
but we also need to show that
$g_+(z,z')$ and $e^{2\lambda^2\phi(z)}g_-(z,z')e^{2\lambda^2\phi(z')}$
have the same asymptotics at the fluxes.
Let us examine $g_\pm(z,z')$
in the neighbourhood of $z=\omega$.
One can see that $g_+(\omega,z')$ is well defined for
$\lambda^2\tilde{\cal V}<2\pi$, while $g_-(\omega,z')=0$.
This does not contradict (\ref{thingtobeproved})
since the exponential 
 $e^{2\lambda^2\phi(z)}$
diverges as
  $\kappa |\mbox{z}-
\mbox{w}|^{-\lambda^2{\tilde{\cal V}}/(2\pi)}$
for $z\sim\omega$ where $\kappa$ is a constant.
Consistency requires that $g_-(z,z')\sim \kappa^{-1}
|\mbox{z}-\mbox{w}|^{\lambda^2\tilde{\cal V}/(2\pi)}
 g_+(\omega,z')
e^{-2\lambda^2\phi(z')}$
for $z\sim\omega$.
One can show that $g_-(z,z')$ decays as it should
in the limit $z\rightarrow\omega$ by considering the derivative
of $g_-(z,z')$:
\begin{eqnarray}\label{derivative}
\!\!\left(-i\partial_{\bar{\mbox{z}}}-\h\mbox{x}_{||}
\right)g_-(z,z')&=&\!\h e^{-2
\lambda^2\phi(z)}
G_+(z-z')\\ \nonumber
&&\!-\frac{e^{-2\lambda^2\phi(z)}}{2\rho_f}G_+(-\omega+z)\int_{\tilde\T^2}
 d^2y'G_-(\omega-y')
e^{-2\lambda^2\phi(y')}G_+(y'-z').
\end{eqnarray}
In the neighbourhood of $z=\omega$, 
$2\pi G_+(-\omega+z)\sim i/(\bar{\mbox{z}}-\bar{\mbox{w}})$,
and so the second term in (\ref{derivative}) dominates (provided
$z'\neq\pm\omega$).
Integrating yields
\begin{equation}
g_-(z,z')\sim\frac{1}{\lambda^2\tilde{\cal V}\kappa\rho_f}
|\mbox{z}-\mbox{w}|^{\lambda^2\tilde{\cal V}/(2\pi)}
\int_{\tilde \T^2}
 d^2y'G_-(\omega-y')e^{-2\lambda^2\phi(y')}
G_+(y'-z'),
\end{equation}
which indeed decays correctly.
Full agreement with (\ref{thingtobeproved})
requires 
\begin{equation}\label{strangeformula}
g_+(\omega,z')=
\frac{e^{2\lambda^2\phi(z')}
}{\lambda^2\tilde{\cal V}\rho_f}
\int_{\tilde \T^2}d^2y'
G_-(\omega-y')e^{-2\lambda^2\phi(y')}
G_+(y'-z').
\end{equation}
To check this one simply notes that away from $z'=\pm\omega$
the left and right hand sides
are annihilated by the same differential operator, 
$\left(i\partial_{\mbox{z}'}-\h\bar{\mbox{x}}_{||}-2i\lambda^2
\partial_{\mbox{z}'}\phi(z')\right)
\left(i\partial_{\bar{\mbox{z}}'}-\h\mbox{x}_{||}\right)$.
It is simple to also check that they agree 
in the neighbourhoods of $z'=\pm\omega$ which completes the proof.

\setcounter{equation}{0}
\section{Two instanton singularities}\label{D}

Consider the 2-component row vectors $v_\pm=(1,\,\,\pm\hat Q)$
which are (formally) eigenvectors of $\Psi(0)$ in that
$v_\pm\Psi(0)=e^{\pm\h|Q|\psi(0)}v_\pm$. We now make the decomposition
$(q_1,\,\,q_2)=\alpha_+ v_+ +\alpha_- v_-$
where the quaternions $\alpha_\pm$ 
are not completely free since
$q_1^*q_2-q_2^*q_1=Q$.
The integrand in the definition of $\rho$ is
\begin{eqnarray}\label{ndaggern}
\tv^{-1}n(z){n^\dagger}(z)&=&
|\alpha_+|^2e^{|Q|\psi(0)}\left[
\CG_+(-z){\CG_+^*(z)}e^{-|Q|\psi(z)}+
\CG_-(-z){\CG_-^*(z)}e^{|Q|\psi(z)}\right]\\ \nonumber 
&&+|\alpha_-|^2e^{-|Q|\psi(0)}\left[
\CG_+(-z){\CG_+^*(z)}e^{|Q|\psi(z)}+
\CG_-(-z){\CG_-^*(z)}e^{-|Q|\psi(z)}\right]\\ \nonumber
&&+\hbox{terms linear in $\alpha_+\alpha_-^*$
and $\alpha_-\alpha_+^*$,}  
\end{eqnarray}
where we have employed the notation
\begin{equation}
\CG_\pm(z)=G_1(z)\pm G_2(z),\end{equation}
not to be confused with the $G_\pm(z)$
introduced in section~\ref{6}!
First, let us consider the singularity structure of the free
Green's functions $\CG_\pm(z)$ which satisfy
$
(-i d_z-x)\CG_\pm(z)=\delta_1(z)\pm\delta_2(z).$
Now $\delta_1^2(z)$ and $\delta_2^2(z)$ are zero except
for all dual lattice points ($z\in\tilde\Lambda$).
However $\delta_1^2(z)+\delta_2^2(z)$ is only singular
at half of the lattice points, while $\delta_1^2(z)-\delta_2^2(z)$
is singular at the remaining dual lattice points.
This can be seen from the following identities
\begin{equation}
\delta_1^2(z)+\delta_2^2(z)=2\cos\left(\h(y_1-y_2)\cdot z
\right)\delta^2(z),\quad
\delta_1^2(z)-\delta_2^2(z)=2i\sin\left(\h(y_1-y_2)\cdot z
\right)\delta^2(z).
\end{equation}
Now since $2(y_1-y_2)\in \Lambda$ it follows that
$\h(y_1-y_2)\cdot z=\h \pi n,\quad n\in\Z$ for $z\in\tilde\Lambda$
which means that either the sine or the cosine must be zero
for $z\in\tilde\Lambda$.
In particular, we see that unlike $\delta_1^2(z)+\delta_2^2(z)$,
$\delta_1^2(z)-\delta_2^2(z)$ has no singularity at $z=0$.
Thus we conclude that $\CG_-(z)$ has no singularity at $z=0$.
In the neighbourhood of  $z=0$ we have 
\bg \CG_+(-z){\CG_+^*(z)}\propto \frac{1}{|\mbox{z}|^2},\quad
\CG_- (-z){\CG_-^*(z)} \hbox{ non-singular.}\eg
We  also require the behaviour of $\psi(z)$ at
$z=0$,
$
\psi(z)\sim -({\tv}/{2\pi}) \log |\mbox{z}|$.
Near $z=0$ we have
\bg\label{singularities}
\CG_+(-z){\CG_+^*(z)}e^{-|Q|\psi(z)}\propto
|\mbox{z}|^{-2+|Q|\tv/(2 \pi)},\quad
\CG_+(-z){\CG_+^*(z)}e^{|Q|\psi(z)}\propto
|\mbox{z}|^{-2-|Q|\tv/(2\pi)}.\\
\eg
The second part of (\ref{singularities}), i.e. 
$\CG_+(-z){\CG_+^*(z)}e^{|Q|\psi(z)}$ is non-integrable.
However, this term is absent in the $|\alpha_+|^2$
contribution to \eq{ndaggern} and so if we make
the choice $\alpha_-=0$ we do not encounter this singularity.
The first part of (\ref{singularities})
is an integrable singularity for $|Q|>0$.
In fact if we take $|Q|\tv>4\pi$
the singularity disappears. However, then
$\CG_- (-z){\CG_-^*(z)}
e^{|Q|\psi(z)}$ will become non integrable.
Accordingly, for the singularities in (\ref{integ})
to be integrable we require $
\alpha_-=0,$ and $
0<|Q|\tv<4\pi$ which implies (\ref{2con1}) and (\ref{2con2}).

\end{document}